\documentclass[9pt]{elife}

\usepackage{lipsum} 
\usepackage[version=4]{mhchem}
\usepackage{siunitx}
\DeclareSIUnit\Molar{M}

\usepackage{booktabs}
\usepackage{chemarr}
\usepackage{kbordermatrix,blkarray}
\usepackage{xcolor}

\title{Nonequilibrium calcium dynamics optimizes the energetic efficiency of mitochondrial metabolism}

\author[1,2]{Valérie Voorsluijs}
\author[2,3]{Francesco Avanzini}
\author[2,4]{Gianmaria Falasco}
\author[2]{Massimiliano Esposito}
\author[1,5,6]{Alexander Skupin}

\affil[1]{Luxembourg Centre for Systems Biomedicine, University of Luxembourg, 6 avenue du Swing, L-4367 Belvaux, Luxembourg}
\affil[2]{Complex Systems and Statistical Mechanics, Department of Physics and Materials Science, University of Luxembourg, 162A avenue de la Faïencerie, L-1511 Luxembourg, Luxembourg}
\affil[3]{Department of Chemical Sciences, University of Padova, 1 Via F. Marzolo, I-35131 Padova, Italy}
\affil[4]{Department of Physics and Astronomy, University of Padova, 8 Via F. Marzolo, I-35131 Padova, Italy}
\affil[5]{Department of Physics and Materials Science, University of Luxembourg, 162 A avenue de la Faïencerie, L-1511 Luxembourg, Luxembourg}
\affil[6]{Department of Neuroscience, University of California San Diego, 9500 Gilman Drive, 92093 San Diego, CA, USA}

\corr{valerie.voorsluijs@uni.lu}{VV}
\corr{francesco.avanzini@unipd.it}{FA}
\corr{gianmaria.falasco@unipd.it}{GF}
\corr{massimiliano.esposito@uni.lu}{ME}
\corr{alexander.skupin@uni.lu}{AS}

\begin{document}

\maketitle

\begin{abstract}
Living organisms continuously harness energy for their survival while part of that energy is dissipated, and determining the efficiency of specific cellular processes remains a largely open problem. Here, we analyze the efficiency of ATP production through the Krebs cycle and oxidative phosphorylation, which generate most of the chemical energy in eukaryotes. The regulation of this pathway by calcium signaling can affect its energetic output, but the concrete energetic impact of this crosstalk remains elusive. Calcium enhances ATP production by activating key enzymes of the Krebs cycle while calcium homeostasis is ATP-dependent. We propose a detailed kinetic model describing the calcium-mitochondria crosstalk and analyze it using nonequilibrium thermodynamics: after identifying the effective reactions driving mitochondrial metabolism out of equilibrium, we quantify the thermodynamic efficiency of mitochondrial metabolism for different physiological conditions. Calcium oscillations boost the efficiency close to substrate-limited conditions, suggesting a compensatory role of calcium signaling in mitochondrial bioenergetics.
\end{abstract}



\section{Introduction}
\label{sec:intro}
Life relies on permanent conversions between different forms of energy, a phenomenon referred to as energy transduction. A wide range of cellular processes are fueled by the chemical energy stored in adenosine triphosphate (ATP), but the compartmentalization of eukaryotic cells also enables the storage of potential energy across the membranes of organelles \citep{nicholls_bioenergetics_1992}. Energy transduction is mediated by enzymes and pumps driven in a nonequilibrium thermodynamic manner by the hydrolysis of ATP, chemical gradients or membrane potentials.

In optimal scenarios where transduction is fully efficient, the input energy is completely transformed into usable work. However, biological processes are typically accompanied by entropy production, \textit{i.e.}, dissipation of energy in the form of heat and/or chemical waste that is unusable for transduction~\citep{calisto_mechanisms_2021}. For example, the action of many transmembrane ionic pumps transporting ions against their concentration gradient is often based on catalysing the hydrolysis of ATP. The chemical energy released by hydrolysis is partly used to drive ionic transport while another part is dissipated. In the extreme case of pump uncoupling, also known as ``slippage'', all the energy of ATP hydrolysis is dissipated without any ion transport~\citep{berman_slippage_2001}.

Different nonequilibrium kinetic models have been developed to account for energy loss in pumps \citep{graber_bioenergetics_1997, rubi_energy_2007, hill_free_2012, wikstrom_thermodynamic_2020} but have only provided limited insights into energetic costs at the pathway level. New approaches based on metabolic network reconstruction and nonequilibrium thermodynamics are gradually emerging to rationalize the energetic costs of cellular processes~\citep{yang_physical_2021} including gene regulation \citep{estrada_information_2016}, repair mechanisms \citep{sartori_thermodynamics_2015, goloubinoff_chaperones_2018}, enzymatic catalysis \citep{flamholz_glycolytic_2013}, information processing \citep{parrondo_thermodynamics_2015} or signaling \citep{cao_free-energy_2015, rodenfels_heat_2019}.
A framework to study energy transduction in complex open chemical reaction networks (CRN) has recently been proposed and used to study the efficiency of pathways of the central energy metabolism in the absence of regulations \citep{wachtel_free-energy_2022}. Evaluating the efficiency of tightly coupled transduction processes, \textit{i.e.} processes whose input and output currents are equal, is straightforward as it does not depend on the net reaction flux. However, when regulations come into play, this tight coupling can be lost and kinetic models become indispensable to evaluate the flux of the different processes contributing to the efficiency.

\begin{figure}
    \centering
    \includegraphics[width=\textwidth]{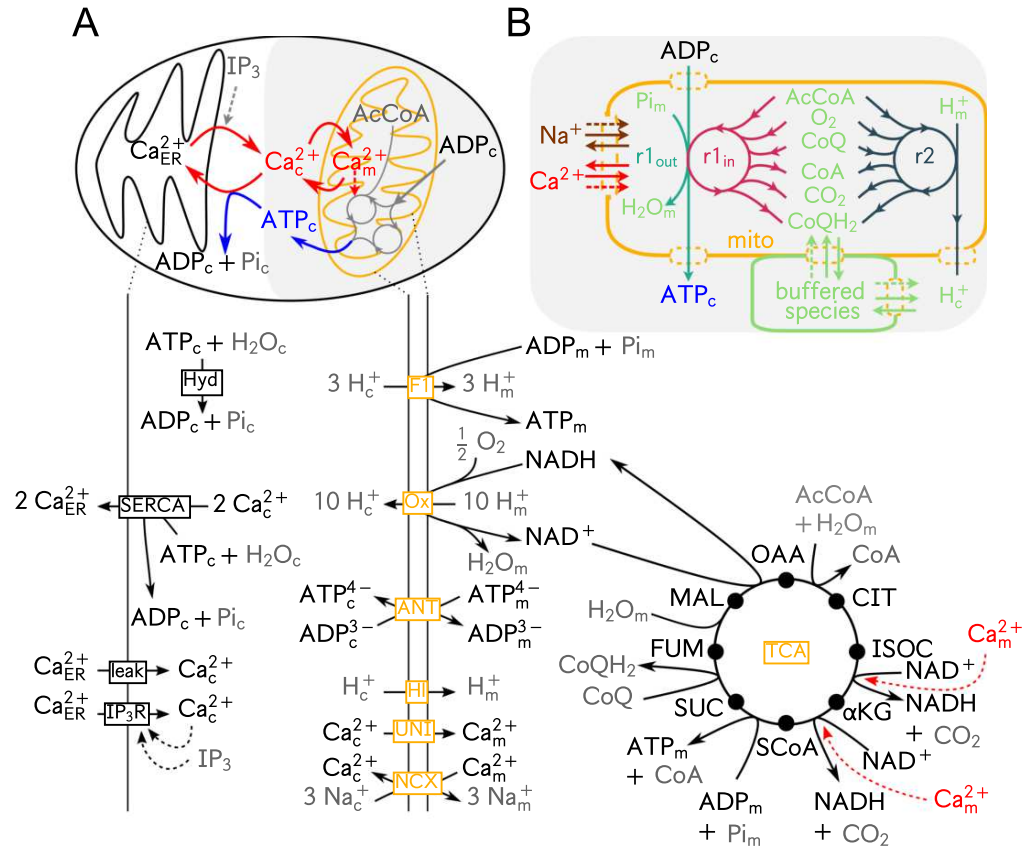}
    \caption{Representation of the model components, conceptualization of mitochondria as a chemical engine and corresponding abbreviations. Balanced chemical equations, detailed expressions of the reaction rates, thermodynamic forces and reference parameter values are given in \TABLE{chemEqn}, \TABLE{fluxes}, \TABLE{forces} and \TABLE{param} in \nameref{sec:methods}, respectively. (\textit{A}) The upper part depicts the \ce{Ca^{2+}} (red) and ATP (blue) fluxes responsible for the crosstalk between \ce{Ca^{2+}} dynamics and mitochondrial metabolism. The bottom part is a detailed description of the model components. The kinetic rates for TCA cycle fluxes and processes involving exchanges across the mitochondrial membrane are respectively originating from Dudycha \citep{dudycha_detailed_2000} and Magnus-Keizer models \citep{magnus_minimal_1997, magnus_model_1998-c, magnus_model_1998-m}, except for the transformation of MAL into OAA, which is described more realistically by a reversible flux \citep{berndt_physiology-based_2015}. Here, OXPHOS corresponds to the net redox reaction resulting from the electron transport chain (Ox) and the synthesis of ATP by the F1F0-ATPase (F1). A last module, consisting of \ce{Ca^{2+}} exchanges across the ER membrane and cytosolic ATP hydrolysis are taken from the models from Komin \textit{et al.} \citep{komin_multiscale_2015} and from Wacquier \textit{et al.} \citep{wacquier_interplay_2016}. Controlled species (\textit{i.e.}, species whose concentration is assumed to be constant) are shown in gray, dynamical species in black and dashed arrows represent regulations. Processes are annotated in yellow and black boxes for mitochondrial and cytosolic/ER processes, respectively. (\textit{B}) Mitochondrial metabolism is conceptualized as an open chemical engine that transforms ADP\textsubscript{c} into ATP\textsubscript{c} through a set of 2 emergent cycles split in 3 effective reactions ($\mathrm{r1_{out}}$, $\mathrm{r1_{in}}$ and $\mathrm{r2}$). Some of the controlled species involved in the internal reactions are buffered at a constant concentrations (green), while \ce{Na^{+}} (brown) and \ce{Ca^{2+}} (red) regulate reaction rates by activating specific enzymes or acting on the mitochondrial membrane potential.
    Abbreviations: AcCoA -- acetyl coenzyme A, $\alpha$KG -- alpha-ketoglutarate, ATP -- Adenosine triphosphate, ADP -- Adenosine diphosphate, CIT -- citrate, CoA -- coenzyme A, CoQ/\ce{COQH2} -- coenzyme Q10, FUM -- fumarate, \ce{IP3} -- inositol 1,4,5-trisphosphate, ISOC -- isocitrate, MAL -- malate, \ce{NAD^{+}}/NADH -- nicotinamide adenine dinucleotide, OAA -- oxaloacetate, Pi -- inorganic phosphate, SUC -- succinate, SCoA -- succinyl coenzyme A.}
    \label{fig:fluxes}
\end{figure}

Here, we resort to a such a kinetically-detailed nonequilibrium thermodynamic approach to show and quantify how active signaling can have a beneficial energetic impact on metabolism. In particular, we analyze the efficiency of the mitochondrial production of ATP \textit{via} the tricarboxylic acid (TCA) cycle and oxidative phosphorylation (OXPHOS), and take into account its regulation by calcium (\ce{Ca^{2+}}). In mitochondria, \ce{Ca^{2+}} activates two key enzymes of the TCA cycle (isocitrate dehydrogenase and $\alpha$-ketoglutarate dehydrogenase)~\citep{mccormack_characterization_1985, hajnoczky_decoding_1995, griffiths_mitochondrial_2009, denton_regulation_2009} and thereby increases the flux of high energy electrons, in the form of NADH, feeding the electron transport chain. The successive redox reactions in the mitochondrial membrane contribute to the establishment of the proton motive force driving the mitochondrial synthesis of ATP by F1F0-ATPase. Depending on the concentration of cytosolic ATP, \ce{Ca^{2+}} can, however, be sequestrated into cell compartments other than mitochondria, such as the endoplasmic reticulum (ER) \textit{via} the sarcoendoplasmic reticulum \ce{Ca^{2+}} ATPase (SERCA), or extruded to the extracellular space~\citep{berridge_calcium--life_1998}. These mechanisms ensure that \ce{Ca^{2+}} does not accumulate in the cytosol, as a persistent high cytosolic \ce{Ca^{2+}} concentration is toxic for the cell. The central coupling enabling the \ce{Ca^{2+}}-mitochondria crosstalk is thus given by the \ce{Ca^{2+}} fluxes between the cytosol and the ER or mitochondria (\FIG{fluxes}). The \ce{Ca^{2+}} release from the ER, by leakage or \textit{via} channels (\ce{IP3}Rs) upon stimulation by inositol 1,4,5-trisphosphate (\ce{IP3}), and \ce{Ca^{2+}} exchanges with mitochondria are ATP-independent, as opposed to \ce{Ca^{2+}} transport into the ER that relies on ATP-consuming SERCA pumps.

Since intracellular \ce{Ca^{2+}} dynamics is strongly nonlinear (which can lead to oscillations in \ce{Ca^{2+}} concentration) and depends itself on ATP availability, evaluating the net effect of signaling on the energetic efficiency of mitochondrial metabolism is not straightforward. Our analysis quantifies the energetic efficiency of this essential cellular process beyond steady-state conditions, such as in an oscillatory regime. Overall, the proposed framework is laying the foundations for a more comprehensive characterization of energetic costs in biology.


\section{Results}
\label{sec:results}

\subsection{Modeling and theoretical frameworks}
We developed a curated model for the essential \ce{Ca^{2+}}-metabolism system integrating different modules~\citep{magnus_minimal_1997, magnus_model_1998-c, magnus_model_1998-m, dudycha_detailed_2000, cortassa_integrated_2003, bertram_simplified_2006, wei_mitochondrial_2011, komin_multiscale_2015, berndt_physiology-based_2015, wacquier_interplay_2016}. Its comprehensive parameterization on experimental data represent a major step towards the detailed analysis of the mitochondrial regulation by \ce{Ca^{2+}}. The underlying kinetic models originally aimed at capturing the essential mechanisms of the the \ce{Ca^{2+}}-metabolism interplay and at rationalizing experimental data about the response of \ce{Ca^{2+}} signals to changes in mitochondrial activity (and \textit{vice versa}). After refining these models to combine them in a coherent way, 
we analyzed the coupled pathways by a nonequilibrium thermodynamic description of CRN \citep{rao_nonequilibrium_2016, rao_conservation_2018, wachtel_thermodynamically_2018, avanzini_thermodynamics_2020, avanzini_nonequilibrium_2021, avanzini_thermodynamics_2022, avanzini_circuit_2023, wachtel_free-energy_2022}.

To compute their metabolic efficiency, we analyzed mitochondria as out-of-equilibrium chemical engines (\FIG{fluxes}\textit{B}) satisfying the second law of thermodynamics~\citep{rao_conservation_2018, avanzini_nonequilibrium_2021}:
\begin{equation}
    T \sigma = - \mathrm{d}_t \mathcal{G} + \dot{w}_\mathrm{nc} + \dot{w}_\mathrm{driv}\,.\label{eq:2law_MAIN}
\end{equation}
Mitochondrial metabolism constitutes an open CRN that continuously harnesses the free energy stored in buffered species (e.g., AcCoA, CoQ, \ce{O2}, $\mathrm{H}^{+}_\mathrm{m}$) to synthesize ATP\textsubscript{c} from ADP\textsubscript{c} while being influenced by \ce{Na^{+}} homeostasis and cytosolic processes such as \ce{Ca^{2+}} signaling and ATP\textsubscript{c} consumption. From a thermodynamic perspective, the synthesis of ATP\textsubscript{c} and the regulations correspond to free energy exchanges between the mitochondrial engine and its surroundings. They appear in the second law (\EQ{2law_MAIN}) as the nonconservative work rate, $\dot{w}_\mathrm{nc}$, and the driving work rate, $\dot{w}_\mathrm{driv}$, respectively. $\dot{w}_\mathrm{nc}$ is the energy current maintaining the CRN out of equilibrium while $\dot{w}_\mathrm{driv}$ is the energy current resulting from the modification of the underlying equilibrium state by the out-of-equilibrium dynamics (\FIG{thermo}\textit{B}). The difference between their sum and the variation in time of the internal Gibbs free energy of mitochondria, $\mathcal{G}$, equals the free energy dissipated by the mitochondrial reactions, \textit{i.e.} the entropy production rate (EPR) $\sigma$ times the absolute temperature~$T$.

The expressions of the thermodynamic quantities in \EQ{2law_MAIN} are derived for mitochondrial metabolism using a topological analysis (developed in~\cite{avanzini_thermodynamics_2020, avanzini_nonequilibrium_2021, avanzini_circuit_2023}) of the corresponding CRN, which allowed us to identify conservation laws and emergent cycles. The conservation laws define parts of molecules that remain intact in all mitochondrial reactions and are instrumental to determine the Gibbs free energy~$\mathcal{G}$. The emergent cycles define the 3 effective reactions 
    \begin{align}
        \mathrm{ADP}_\mathrm{c} + \mathrm{Pi}_\mathrm{m} & \xrightleftharpoons[]{\mathrm{\mathbf{r1_{out}}}} \mathrm{ATP}_\mathrm{c} + \mathrm{H}_2 \mathrm{O}_\mathrm{m} \label{eq:chemOutput_MAIN}\\
        \frac{3}{22}\, \mathrm{O}_2 + \frac{1}{11}\, \mathrm{AcCoA} + \frac{1}{11}\, \mathrm{CoQ} & \xrightleftharpoons[]{\mathrm{\mathbf{r1_{in}}}} \frac{1}{11}\, \mathrm{CoA} + \frac{2}{11}\, \mathrm{CO}_2 + \frac{1}{11}\, \mathrm{CoQH}_2 \label{eq:chemInputTCA_MAIN}\\
      \mathrm{H}^{+}_\mathrm{m} +  \frac{1}{22}\, \mathrm{O}_2 + \frac{1}{33}\, \mathrm{AcCoA} + \frac{1}{33}\, \mathrm{CoQ} & \xrightleftharpoons[]{\mathbf{r2}} \mathrm{H}^{+}_\mathrm{c} + \frac{1}{33}\, \mathrm{CoA} + \frac{2}{33}\, \mathrm{CO}_2 + \frac{1}{33}\, \mathrm{CoQH}_2. \label{eq:emHc_MAIN}
    \end{align}
and split the nonconservative work rate into the sum of 3 contributions: $\dot{w}_\mathrm{nc} = \dot{w}_\mathrm{r1_{out}} + \dot{w}_\mathrm{r1_{in}}+ \dot{w}_\mathrm{r2}$, where $\dot{w}_\mathrm{r1_{out}}$ quantifies the mitochondrial free energy output corresponding to the synthesis of ATP in the cytosol~(\EQ{chemOutput_MAIN}), while $\dot{w}_\mathrm{r1_{in}}$ and $\dot{w}_\mathrm{r2}$ quantify the mitochondrial free energy power source due the interconversion of the buffered species \textit{via} reactions in \EQ{chemInputTCA_MAIN} and~\EQ{emHc_MAIN}, respectively.

The average thermodynamic efficiency $\bar{\eta}$ of mitochondria
can then be calculated as
\begin{equation}
    \bar{\eta} = - \frac{\bar{w}_\mathrm{nc}^\mathrm{output}}{\bar{w}_\mathrm{nc}^\mathrm{input} + \bar{w}_\mathrm{driv}}
\end{equation}
where $\bar{w}_\mathrm{nc}^\mathrm{input}=\bar{w}_\mathrm{r1_{in}}+\bar{w}_\mathrm{r2}$ and $\bar{w}_\mathrm{nc}^\mathrm{output} = \bar{w}_\mathrm{r1_{out}}$,
and the overline denotes either steady-state quantities or averages over one period of \ce{Ca^{2+}} oscillations (notice that $\overline{\mathrm{d}_t \mathcal{G}} = 0$).

The EPR nonconservative and driving work contributions vanish at equilibrium according to the second law of thermodynamics but take finite values in nonequilibrium regimes (\FIG{thermo}\textit{B}). 
The nonequilibrium kinetics of the system was assessed for different stimulation conditions and mitochondrial substrate concentrations (\textit{i.e.} for different $\left[\mathrm{IP_3}\right]$ and $\left[\mathrm{AcCoA}\right]$ in the simulations), which allowed for the calculation of the corresponding nonconservative and driving work contributions and, ultimately, of the efficiency of mitochondrial metabolism.


\subsection{\ce{Ca^{2+}}-metabolism crosstalk affects the oscillation period and the production of ATP}

\begin{figure}
    \centering
    \includegraphics[width=0.7\linewidth]{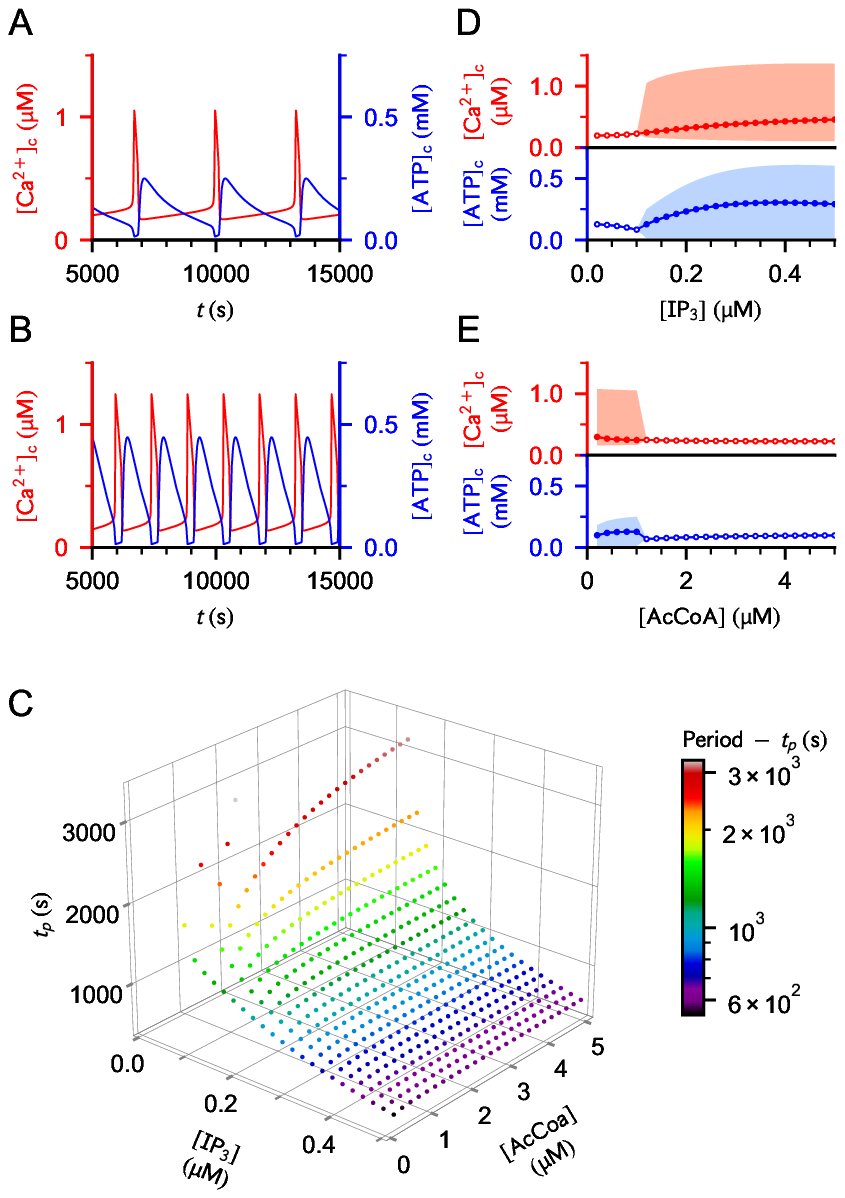}
    \caption{Kinetic behavior of the system. (\textit{A}-\textit{B}) $\mathrm{Ca}^{2+}_\mathrm{c}$ and ATP\textsubscript{c} concentrations over time for $\left[\mathrm{AcCoA}\right]=1\,\mathrm{\mu M}$ and (\textit{A}) $\left[\mathrm{IP_3}\right]=0.12\,\mathrm{\mu M}$ or (\textit{B}) $\left[\mathrm{IP_3}\right]=0.20\,\mathrm{\mu M}$. (\textit{C}) Effect of $\left[\mathrm{IP_3}\right]$ and $\left[\mathrm{AcCoA}\right]$ on the oscillation period. (\textit{D}-\textit{E}) Average concentration of $\mathrm{Ca}^{2+}_\mathrm{c}$ and ATP\textsubscript{c} as a function of (\textit{D}) $\left[\mathrm{IP_3}\right]$ for $\left[\mathrm{AcCoA}\right]=1\,\mathrm{\mu M}$ or as a function of (\textit{E}) $\left[\mathrm{AcCoA}\right]$ for $\left[\mathrm{IP_3}\right]=0.12\,\mathrm{\mu M}$. Empty and filled dots represent steady-state and oscillatory regimes, respectively, and the boundaries of the shaded areas correspond to the minimum and maximum concentrations. Parameter values are given in \TABLE{param}. \FIGSUPP[thermo]{thermo3DSupp}B illustrates the behavior of $\left[\mathrm{ATP}\right]_\mathrm{c}$ for an extended range of $\left[\mathrm{AcCoA}\right]$ and $\left[\mathrm{IP_3}\right]$.
    }
    \label{fig:kinetics}
    \figsupp{Average SERCA, UNI and Ca\textsuperscript{2+}-dependent TCA fluxes and average cytosolic and mitochondrial Ca\textsuperscript{2+} concentrations \textit{vs.} $\left[\mathrm{IP_3}\right]$ as complementary figures to the bifurcation diagrams shown in \FIG{kinetics}\textit{D} and \FIG{thermo}\textit{C} bottom and portrait phases in \FIG{uncoupled}\textit{C}. Note that $\bar{J}_\mathrm{IDH}$ and $\bar{J}_\mathrm{KGDH}$ are indistinguishable. Empty and filled dots correspond to steady-state or period-averaged quantities, respectively. Parameter values are the same as in \FIG{kinetics}\textit{D}.}{\includegraphics[width=0.8\textwidth]{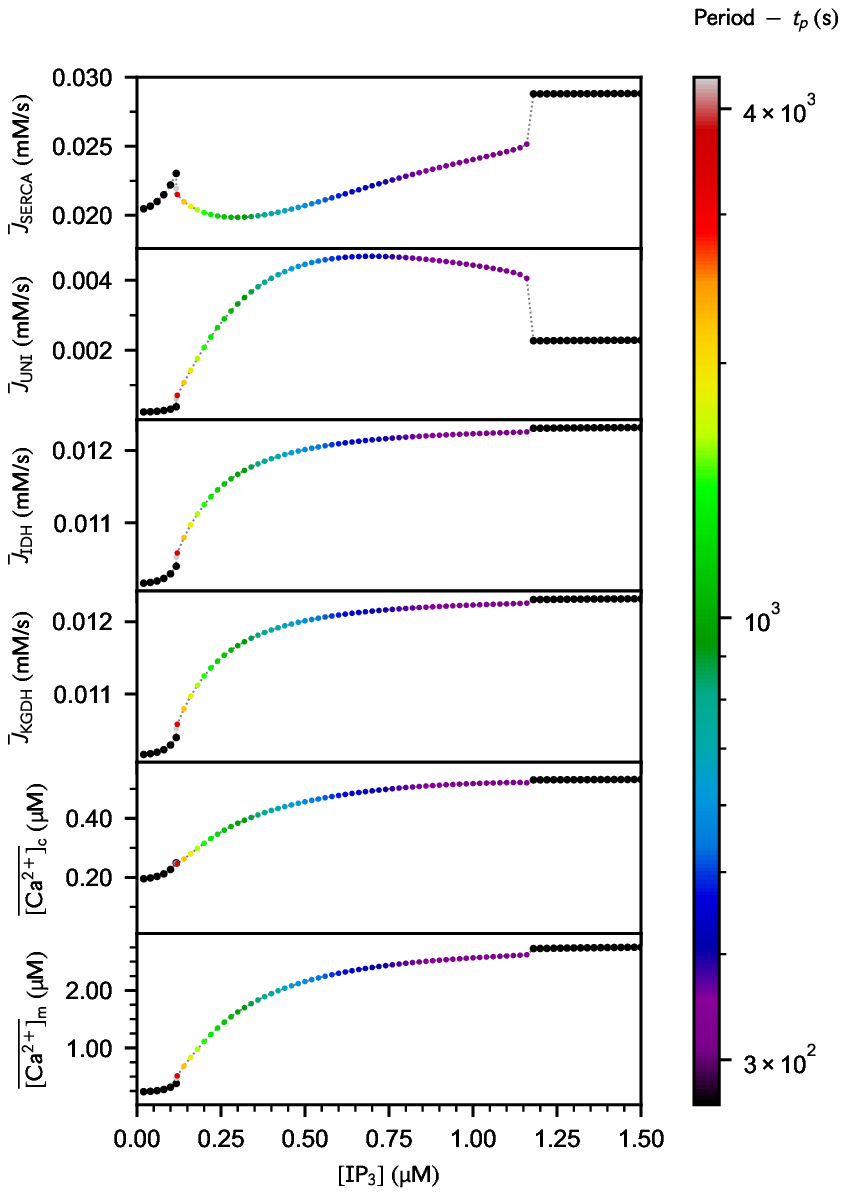}}\label{figsupp:kineticsSupp}
\end{figure}

To validate the kinetic model, we compared our simulation results to experimental and simulation data from the literature. Slow spiking is found around the bifurcation point corresponding to the transition from steady-state to oscillations, which marks the onset of the signaling machinery. A decrease in the oscillation period is observed as $\left[\mathrm{IP}_3\right]$ is increased (\FIG{kinetics}\textit{A}-\textit{C}) or as $\left[\mathrm{AcCoA}\right]$ is decreased (\FIG{kinetics}\textit{C}). These trends are in agreement with stimulation experiments performed in various cell types \citep{woods_repetitive_1986, falcke_reading_2004, dupont_calcium_2007, thurley_reliable_2014, moein_dissecting_2017} and with behaviors reported for limited availability of mitochondrial substrate \citep{jouaville_synchronization_1995, wacquier_interplay_2016, moein_dissecting_2017}.
In the oscillatory regime, $\left[\mathrm{ATP}\right]_\mathrm{c}$ displays a maximum in dependence on $\left[\mathrm{IP}_3\right]$ and $\left[\mathrm{AcCoA}\right]$ (\FIG{kinetics}\textit{D-E} and \FIGSUPP[thermo]{thermoSupp}), a feature that is also predicted by the model of Wacquier \textit{et al.}~\citep{wacquier_interplay_2016}. In our simulations, a cusp in the average of $\left[\mathrm{ATP}\right]_\mathrm{c}$ is additionally observed at the critical point (\FIG{kinetics}\textit{D-E}).

Most of these observations can be rationalized based on the dependence of SERCA pumps on ATP\textsubscript{c}, which enables the switch between ER and mitochondrial \ce{Ca^{2+}} sequestration and is a key signature of the \ce{Ca^{2+}}-metabolism crosstalk. As $\left[\mathrm{IP}_3\right]$ increases, more \ce{Ca^{2+}} is released into the cytosol through IP\textsubscript{3}Rs. The steady-state $\left[\mathrm{ATP}\right]_\mathrm{c}$ thus decreases due to a more demanding maintenance of the basal $\left[\mathrm{Ca}^{2+}\right]_\mathrm{c}$ \textit{via} SERCA pumps. At the critical $\left[\mathrm{IP}_3\right]$ corresponding to the onset of oscillations, mitochondrial sequestration of \ce{Ca^{2+}} becomes significant, which not only relieves SERCA pumps but also enables the activation of \ce{Ca^{2+}}-sensitive dehydrogenases of the TCA cycle (\FIGSUPP[kinetics]{kineticsSupp}). These combined effects result in an increase of the average $\left[\mathrm{ATP}\right]_\mathrm{c}$.
Increasing $\left[\mathrm{IP_3}\right]$ further leads to saturation in mitochondrial buffering of \ce{Ca^{2+}} (\FIGSUPP[kinetics]{kineticsSupp}). More intense \ce{Ca^{2+}} sequestration \textit{via} SERCA pumps is then required and the associated ATP\textsubscript{c} consumption is no longer counterbalanced by the \ce{Ca^{2+}}-enhanced mitochondrial activity, which results in a slow decrease in average $\left[\mathrm{ATP}\right]_\mathrm{c}$.
Meanwhile, increasing stimulation by IP\textsubscript{3} favors more frequent opening of the IP\textsubscript{3}Rs, which results in an decrease of the oscillation period.
In most mathematical models for \ce{Ca^{2+}} signaling and in agreement with experimental observations, the oscillation period saturates at high $\left[\mathrm{IP_3}\right]$ \citep{eisner_study_1986} and, beyond a critical $\left[\mathrm{IP_3}\right]$, oscillations disappear. The cell then exhibits a high-$\left[\mathrm{Ca^{2+}}\right]_\mathrm{c}$ steady-state \citep{falcke_reading_2004}, as reproduced by our simulations (\FIGSUPP[kinetics]{kineticsSupp}). Further stimulation by IP\textsubscript{3} does not affect the steady-state concentrations reached after termination of the oscillations (see \FIG{thermo}\textit{C} bottom for $\left[\mathrm{ATP}\right]_\mathrm{c}$ and \FIGSUPP[kinetics]{kineticsSupp} for $\left[\mathrm{Ca}^{2+}\right]_\mathrm{c}$ and $\left[\mathrm{Ca}^{2+}\right]_\mathrm{m}$), suggesting that IP\textsubscript{3}Rs have reached their maximal release rate and contribute to saturation effect.

The impact of AcCoA level on $\left[\mathrm{ATP}\right]_\mathrm{c}$ and oscillation period is more visible in starving conditions, \textit{i.e.} $\left[\mathrm{AcCoA}\right] \leq 1~\mathrm{\mu M}$, and for low stimulation by IP\textsubscript{3}, \textit{i.e.} $\left[\mathrm{IP}_3\right] \leq 0.24~\mathrm{\mu M}$. ATP production decreases as $\left[\mathrm{AcCoA}\right]$ is decreased except at the onset of oscillations, where a cusp in $\left[\mathrm{ATP}\right]_\mathrm{c}$ occurs (\FIG{kinetics}\textit{E} and \FIGSUPP[kinetics]{kineticsSupp}\textit{B}). At this critical point, $\left[\mathrm{ATP}\right]_\mathrm{c}$ has become limiting for \ce{Ca^{2+}} uptake by SERCAs and mitochondrial exchanges take over. This switch allows for activation of TCA cycle enzymes by \ce{Ca^{2+}} and locally rescues ATP production. Mitochondrial exchanges intensify as $\left[\mathrm{AcCoA}\right]$ is further decreased. The larger \ce{Ca^{2+}} efflux from mitochondria exerts a positive feedback on \ce{IP3}Rs which open more frequently, hence the decrease in oscillation period.


\subsection{The efficiency of mitochondrial metabolism displays a maximum in the regime of \ce{Ca^{2+}} spiking}

\begin{figure}
    \centering
    \includegraphics[width=\linewidth]{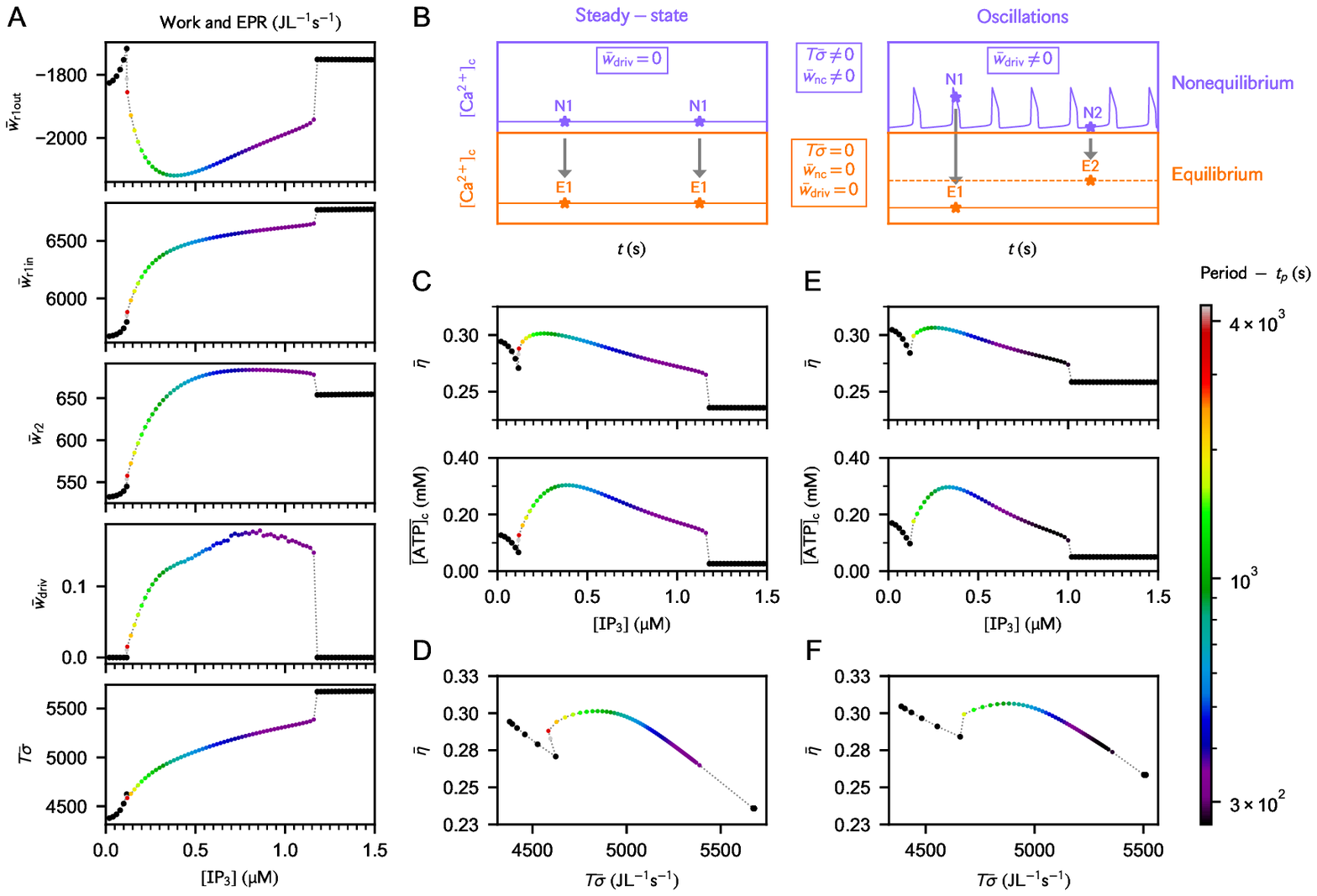}
    \caption{Stimulating the Ca\textsuperscript{2+} signaling machinery impacts the dissipation and efficiency of mitochondrial metabolism \textit{via} the Ca\textsuperscript{2+}-metabolism crosstalk. (\textit{A})~Nonconservative work contributions, driving work and dissipation for different $\left[\mathrm{IP_3}\right]$. The driving work represents less than 0.01\% of the EPR. At high stimulation, oscillations disappear in the favor of a nonequilibrium steady-state regime. (\textit{B})~Expected work contributions in equilibrium and nonequilibrium conditions. Stars denote the state of the system at different time points in nonequilibrium conditions (violet) and the corresponding underlying equilibrium state (orange). As illustrated, the non-zero driving work contribution in the oscillatory regime can modify the underlying equilibrium state of the system. (\textit{C})~Efficiency and ATP\textsubscript{c} concentration as a function of $\left[\mathrm{IP_3}\right]$. (\textit{D)}~Efficiency as a function of the total dissipation for a range of $\left[\mathrm{IP_3}\right]$ extended up to $5\, \mu\mathrm{M}$. (\textit{E}-\textit{F})~Plots corresponding to (\textit{C}-\textit{D})~for $V_{max}^\mathrm{SERCA} = 0.08\, \mathrm{\mu M\, s^{-1}}$. Empty and filled dots correspond to steady-state or period-averaged quantities, respectively. Unless specified otherwise, parameter values are the same as in \FIG{kinetics}\textit{D}. An analogous thermodynamic behavior is observed upon stimulation by AcCoA (\FIGSUPP[thermo]{thermoSupp}).}
    \label{fig:thermo}
    \figsupp{Efficiency of mitochondrial metabolism ($\eta$), average cytosolic ATP concentration ($\overline{\left[\mathrm{ATP}\right]}_\mathrm{c}~(\mathrm{mM})$) and period of Ca\textsuperscript{2+} oscillations ($t_p$) as functions of $\left[\mathrm{AcCoA}\right]$ and $\left[\mathrm{IP_3}\right]$. (\textit{A}-\textit{B}) Summary 3D plots. IP\textsubscript{3} plays a dominant role in the transition between steady-state and oscillations, which usually takes place for concentrations of IP\textsubscript{3} between 0.1 and 0.2 $\mu$M. The onset of oscillations can be triggered for smaller $\left[\mathrm{IP_3}\right]$ in the presence of a very low level of AcCoA (e.g. $\left[\mathrm{AcCoA}\right]=0.2\, \mu\mathrm{M}$), which supports the role of Ca\textsuperscript{2+} oscillations as a rescuing mechanism aiming to improve the efficiency of energy production in stressing situations such as limited access to carbon substrate. For more clarity, representative behaviors of the efficiency (top panels), $\overline{\left[\mathrm{ATP}\right]}_\mathrm{c}~(\mathrm{mM})$ (middle panels) and period (low panels) were plot for (grey lines - \textit{C}) $\left[\mathrm{AcCoA}\right]=1\,\mu\mathrm{M}$ and (brown lines - \textit{D}) $\left[\mathrm{IP_3}\right]=0.12\,\mu\mathrm{M}$, as complementary figures to the bifurcation diagrams shown in Fig.~2\textit{E} and 2\textit{F}, respectively. Note that maxima in efficiency and in $\overline{\left[\mathrm{ATP}\right]}_\mathrm{c}~(\mathrm{mM})$ can also be observed when $\left[\mathrm{AcCoA}\right]$ is varied. Increments in concentrations are of 0.2 $\mu$M (\textit{A}-\textit{B}) or 0.1 $\mu$M (\textit{D}) for AcCoA and of 0.2 $\mu$M (\textit{A} -- \textit{C}) for IP\textsubscript{3}.}{\includegraphics[width=0.8\textwidth]{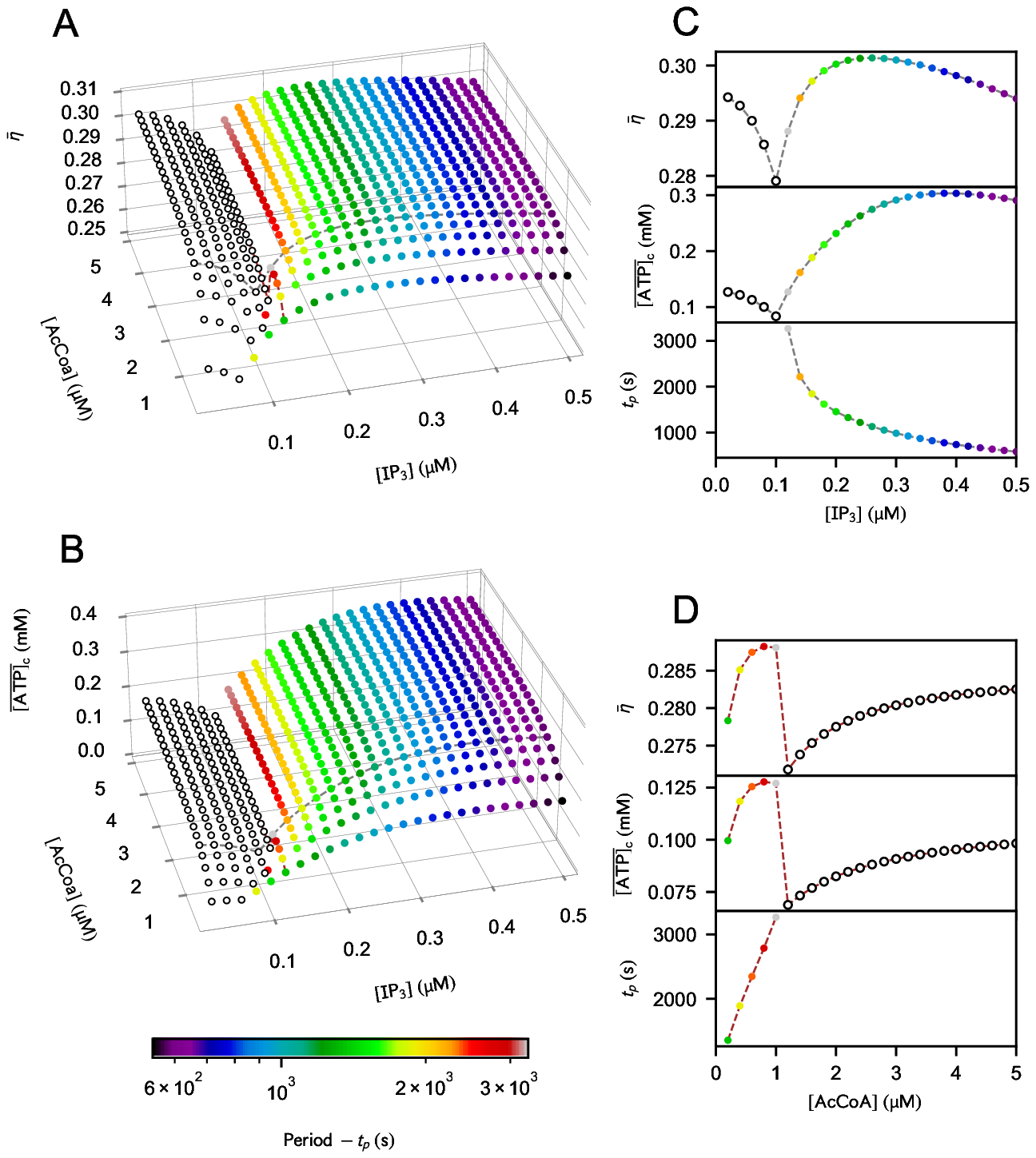}}\label{figsupp:thermo3DSupp}
    \figsupp{Stimulation of mitochondrial metabolism by AcCoA impacts Ca\textsuperscript{2+} dynamics \textit{via} the Ca\textsuperscript{2+}-metabolism crosstalk. (\textit{A})~Nonconservative work contributions, driving work and dissipation for different $\left[\mathrm{AcCoA}\right]$. The driving work represents less than 0.0004\% of the EPR. At high stimulation, oscillations disappear in the favor of a nonequilibrium steady-state regime. (\textit{B})~Efficiency and ATP\textsubscript{c} concentration as a function of $\left[\mathrm{IP_3}\right]$. (\textit{C)}~Efficiency as a function of the total dissipation for the same range of $\left[\mathrm{AcCoA}\right]$ as in (\textit{A}) and (\textit{B}). (\textit{D}-\textit{E})~Plots corresponding to (\textit{B}-\textit{C})~for $V_{max}^\mathrm{SERCA} = 0.08\, \mathrm{\mu M\, s^{-1}}$. Empty and filled dots correspond to steady-state or period-averaged quantities, respectively. Unless specified otherwise, parameter values are the same as in \FIG{kinetics}\textit{D}.}{\includegraphics[width=\linewidth]{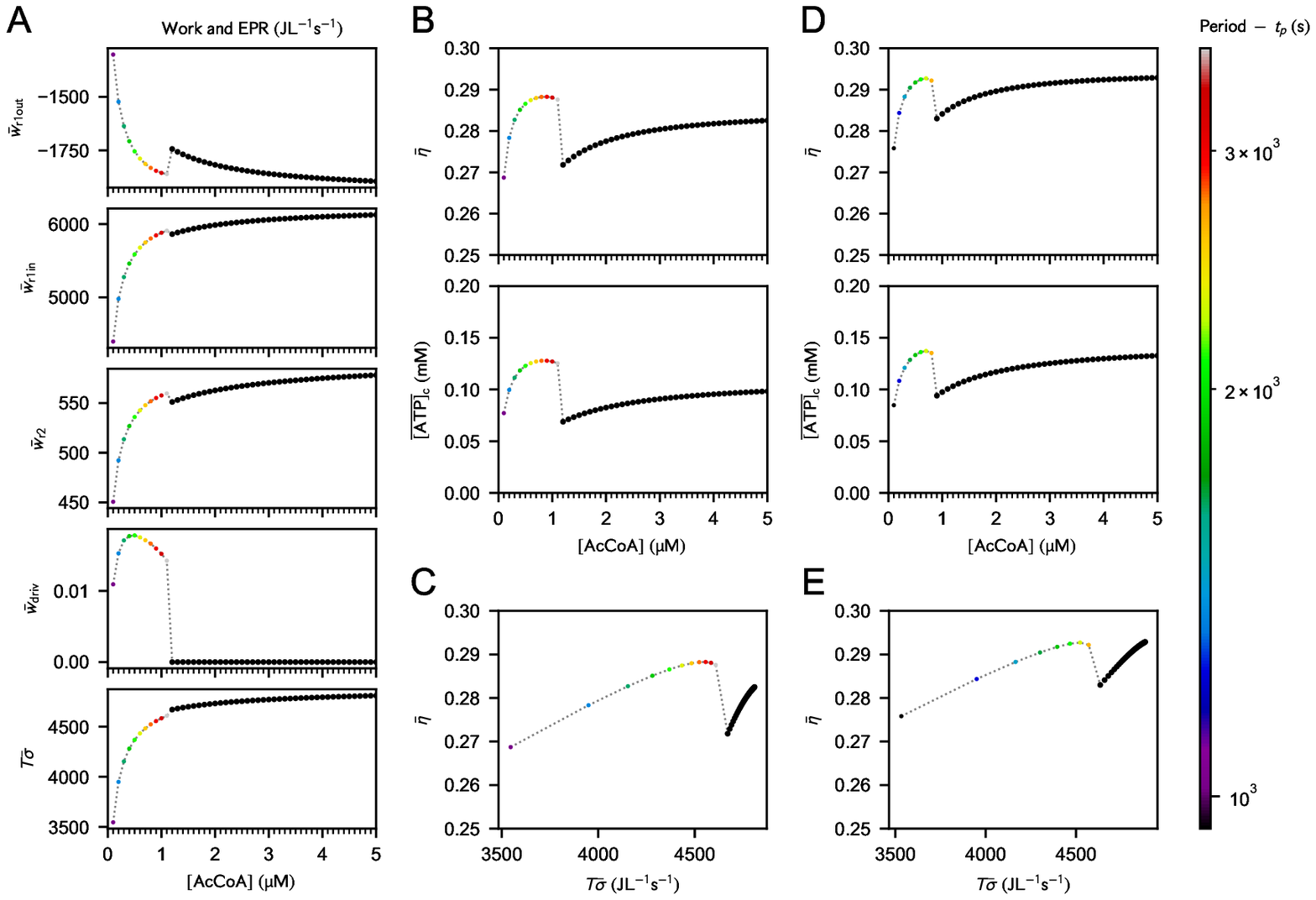}}\label{figsupp:thermoSupp}
\end{figure}

The nonlinear ATP production (\FIG{kinetics}\textit{D-E}) observed for different $\left[\mathrm{IP}_3\right]$ and $\left[\mathrm{AcCoA}\right]$ suggests variations in the output work of mitochondria and, possibly, in the thermodynamic efficiency of their metabolism.
As confirmed computationally, the output nonconservative work ($\bar{w}_\mathrm{r1out}$) displays a minimum (corresponding to maximal \textit{export} of energy from mitochondria) that coincides with the maximal $\left[\mathrm{ATP}\right]_\mathrm{c}$ in the kinetic simulations (\FIG{thermo}\textit{A}~and~\FIGSUPP[thermo]{thermoSupp}\textit{A} top, \textit{vs} \FIG{thermo}\textit{C}~and~\FIGSUPP[thermo]{thermoSupp}\textit{B} bottom). In the extreme case where oscillations disappear for large stimulation by IP\textsubscript{3}, the efficiency drops and reaches a plateau (\FIG{thermo}\textit{E} top). 
On the other hand, both the non-conservative input work contributions ($\bar{w}_\mathrm{r1in}$ and $\bar{w}_\mathrm{r2}$) increase with $[\mathrm{IP}_3]$ and $[\mathrm{AcCoA]}$, while the driving work ($\bar{w}_\mathrm{driv}$) is always negligible compared to the total dissipation (\FIG{thermo}\textit{A}~and~\FIGSUPP[thermo]{thermoSupp}\textit{A}).

Importantly, the maximum in $\left[\mathrm{ATP}\right]_\mathrm{c}$ translates into a maximum in the efficiency of mitochondrial metabolism (\FIG{thermo}\textit{C} and \FIGSUPP[thermo]{thermoSupp}\textit{B}). Such maxima are not systematically observed when $\left[\mathrm{AcCoA}\right]$ is varied at fixed $\left[\mathrm{IP_3}\right]$ (\FIGSUPP[kinetics]{kineticsSupp}\textit{A-B}). However, the increase in efficiency at the onset of the oscillatory regime is a robust feature that points to the stabilizing effect of \ce{Ca^{2+}} spikes on mitochondrial energetics.


\subsection{The relation between mitochondrial efficiency and dissipation is different in fast \ce{Ca^{2+}} spiking regimes triggered by starvation or overstimulation by \ce{IP3}}
Like in other biological processes such as the migration of molecular motors along microtubules, kinetic proofreading or the regulation of circadian clocks \citep{baiesi_life_2018}, the system's efficiency is maximal at intermediate levels of dissipation, corresponding to a limited range of $\left[\mathrm{IP_3}\right]$ and $\left[\mathrm{AcCoA}\right]$ (\FIG{thermo}\textit{E} and \FIGSUPP[thermo]{thermoSupp}\textit{D}).

Overstimulation of the signaling machinery by \ce{IP3} is counterproductive since it only increases dissipation (\FIG{thermo}\textit{A} bottom). By exploring the behavior of efficiency and EPR at large $\left[\mathrm{IP_3}\right]$, we observed a saturation effect (\FIG{thermo}\textit{A}, \textit{C} and \textit{D}) leading to limiting values for the efficiency ($\approx 0.236$) and total dissipation ($\approx 5680\, \mathrm{JL^{-1}s^{-1}}$). The dependency of efficiency on the total dissipation is highly nonlinear (\FIG{thermo}\textit{D}). Around the onset of oscillations (and for a limited range of dissipation rates), a given dissipation rate can be associated to different efficiencies, in which case the highest efficiency is always reached for the highest \ce{Ca^{2+}} spiking frequency, while the lowest efficiency corresponds to the steady-state regime. Reversely, different dissipative regimes can yield the same efficiency. In that case, steady-state regimes display the lowest EPR while the fast-spiking regimes are the more dissipative regimes. We hypothesize that in such instances, the selection of the dissipative regime could be guided by constraints imposed by the global energy budget of the cell.

As $\left[\mathrm{AcCoA}\right]$ is increased, slow-spiking regimes are more dissipative than the fast-spiking ones (\FIGSUPP[thermo]{thermoSupp}\textit{A}, bottom panel) and the efficiency increases almost linearly with dissipation in the oscillatory regime (\FIGSUPP[thermo]{thermoSupp}\textit{C}), which contrasts with the overstimulation by IP\textsubscript{3}. 
A reason for this difference might be that upon overstimulation by IP\textsubscript{3}, ATP\textsubscript{c} hydrolysis overbalances the enhanced production of ATP by mitochondrial metabolism, which is not the case at low $\left[\mathrm{AcCoA}\right]$.

To summarize, fast spiking is less efficient than the slow spiking observed around the bifurcation point, and is also more dissipative when oscillations frequency intensifies due to perturbations in $\left[\mathrm{IP_3}\right]$.


\subsection{Robustness of the efficiency-rescuing effect of \ce{Ca^{2+}} oscillations}

\begin{figure}
    \centering
    \includegraphics[width=0.7\textwidth]{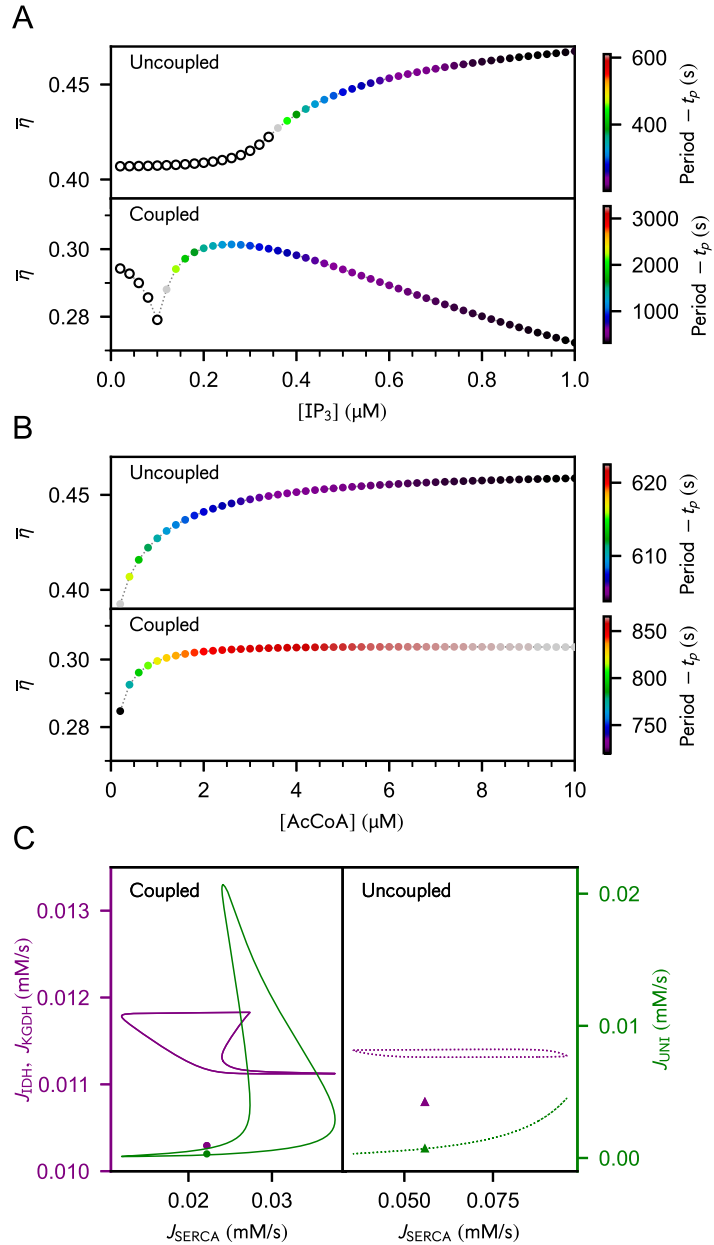}
    \caption{Efficiency of mitochondrial metabolism regulated by Ca\textsuperscript{2+} signaling, Ca\textsuperscript{2+} sequestration fluxes and Ca\textsuperscript{2+}-dependent TCA cycle reaction fluxes, without and with the coupling of SERCA pumps to ATP\textsubscript{c} hydrolysis. (\textit{A}) Stimulating Ca\textsuperscript{2+} release by IP\textsubscript{3}Rs (that is, increasing $\left[\mathrm{IP_3}\right]$) monotonically increases the efficiency in the uncoupled case, which strongly contrasts with the nonmonotonic dependency on $\left[\mathrm{IP_3}\right]$ in the coupled case. While varying over different ranges, the period evolves according to the same trends in both cases. (\textit{B}) Both systems display similar responses in their efficiency upon variations in $\left[\mathrm{AcCoA}\right]$, but the oscillation period of the uncoupled system does not decrease -- and is even slightly increasing -- in stressing conditions corresponding to substrate depletion. As this observation is in contradiction with experimental evidence, the larger efficiencies reached in the uncoupled case are non physiological. Empty and filled dots correspond respectively to steady-state and oscillatory regimes -- note the use of linear colorbar schemes for the period. (\textit{C}) Phase portraits of Ca\textsuperscript{2+}-dependent TCA cycle currents (purple) and mitochondrial Ca\textsuperscript{2+} uptake (green), namely $J_\mathrm{IDH}$, $J_\mathrm{KGDH}$ and $J_\mathrm{UNI}$, \textit{vs.} ER Ca\textsuperscript{2+} uptake, namely $J_\mathrm{SERCA}$. Note that $J_\mathrm{IDH}$ and $J_\mathrm{KGDH}$ are indistinguishable. Symbols denote steady-state values. Triangles ($\left[\mathrm{IP_3}\right]=0.34\,\mathrm{\mu M}$) and dotted curves ($\left[\mathrm{IP_3}\right]=0.42\,\mathrm{\mu M}$) correspond to the uncoupled case, while circles ($\left[\mathrm{IP_3}\right]=0.10\,\mathrm{\mu M}$) and solid curves ($\left[\mathrm{IP_3}\right]=0.24\,\mathrm{\mu M}$) represent the coupled case. (\textit{A} and \textit{C}) $\left[\mathrm{AcCoA}\right]=1\,\mathrm{\mu M}$, (\textit{B}) $\left[\mathrm{IP_3}\right]=0.36\,\mathrm{\mu M}$ and the other parameter values are the same as in \FIG{kinetics}--\FIG{thermo}.}
    \label{fig:uncoupled}
\end{figure}

Our proposed mechanism for the maximum in efficiency relies on the dependence of the SERCA flux ($J_\mathrm{SERCA}$) on the hydrolysis of ATP\textsubscript{c} and on the resulting modulation of \ce{Ca^{2+}} sequestration mechanisms. If \ce{Ca^{2+}} homeostasis was ATP-independent, \ce{Ca^{2+}} would always exert a positive feedback on the TCA cycle flux and the efficiency of metabolism would increase monotonically with the \ce{Ca^{2+}} release from IP\textsubscript{3}Rs. We validated this hypothesis by performing simulations with a modified SERCA flux that is uncoupled from ATP\textsubscript{c} hydrolysis. The degradation of ATP\textsubscript{c} then relies exclusively on other ATP-consuming processes mimicking cellular activity (Hyd reaction in \FIG{fluxes}\textit{A}). As expected, the \ce{Ca^{2+}}-enhanced ATP production by mitochondria is not restrained upon more intense stimulation by \ce{IP3} (\FIG{uncoupled}\textit{A}).
This uncoupling also disables the feedback of ATP production by mitochondria on \ce{Ca^{2+}} oscillations: instead of decreasing, the spike period is barely changed as $\left[\mathrm{AcCoA}\right]$ decreases (\FIG{uncoupled}\textit{B}).

In the uncoupled case, $J_\mathrm{SERCA}$ is not limited by the depletion of ATP\textsubscript{c} and both removal mechanisms proceed synchronously, although \ce{Ca^{2+}} uptake to the ER is predominant (\FIG{uncoupled}\textit{C}, green dotted curve). While the mitochondrial \ce{Ca^{2+}} influx $J_\mathrm{UNI}$ slightly increases with $J_\mathrm{SERCA}$, the \ce{Ca^{2+}}-dependent currents of the TCA cycle, $J_\mathrm{IDH}$ and $J_\mathrm{KGDH}$ (\FIG{uncoupled}\textit{C}, purple dotted curves), are barely affected. Upon regulation by ATP\textsubscript{c}, $J_\mathrm{SERCA}$ varies over a more restricted range, is on average smaller and proceeds with a slight phase shift with respect to $J_\mathrm{UNI}$, which allows for a larger \ce{Ca^{2+}} influx in mitochondria and a more intense activation of the TCA cycle enzymes (\FIG{uncoupled}\textit{C} solid curves).
This mechanism supports the ``efficiency-rescuing'' role of mitochondrial buffering that is visible near the onset of oscillations in the coupled case (\FIG{thermo}\textit{C} and \FIG{uncoupled}\textit{A} bottom), while no such cusp-like transition is observed in the uncoupled case (\FIG{uncoupled}\textit{A} top).

We also explored the robustness of our results against perturbations in the uptake rate of SERCA pumps of the original model. We mimicked the inhibition of SERCA pumps by decreasing the limiting rate $V_{max}^\mathrm{SERCA}$. The efficiency-dissipation relation displays the same features as in the non-inhibited case (\FIG{thermo}\textit{F} and \FIGSUPP[thermo]{thermoSupp}\textit{E} \textit{vs} \FIG{thermo}\textit{D} and \FIGSUPP[thermo]{thermoSupp}\textit{C}, respectively).

Together, these results confirm that the crosstalk between \ce{Ca^{2+}} signaling and mitochondrial energy metabolism is a major mechanism underlying the maximum in efficiency arising in the spiking regime, even when the amplitude of this coupling is reduced due to the inhibition of SERCA pumps.


\section{Discussion}

Here, we examined the impact of \ce{Ca^{2+}} signaling on the efficiency of mitochondrial metabolism by using tools from the nonequilibrium thermodynamics of CRNs on a detailed and experimentally validated kinetic model of the \ce{Ca^{2+}}-metabolism crosstalk. Our results highlight that, despite a usually higher dissipation rate compared to steady-state regimes, \ce{Ca^{2+}} oscillations can enhance the efficiency of mitochondrial metabolism. In particular, stimulation by IP\textsubscript{3} reduces the steady-state efficiency of metabolism but at the onset of oscillations
the efficiency raises with a cusp-like transition and reaches a maximum of about 30\% before decreasing again at higher stimulation. This value corresponds to the efficiency of the TCA cycle estimated in the absence of regulation with a nonequilibrium thermodynamic approach \citep{wachtel_free-energy_2022}. Moreover, slow-spiking is less dissipative than fast-spiking. Thus, we hypothesize that, for a given cell state, there exists an optimal stimulation level leading to slow-spiking/low-dissipation oscillations which maximize the efficiency of metabolism during signaling. For higher stimulation, the \ce{Ca^{2+}} signaling machinery then generates more dissipative regimes of gradually decreasing efficiency.

In the broader context of physical bioenergetics, energetic costs are usually assessed by evaluating the Gibbs free energy of reaction ($\Delta_r G$) dissipated or the equivalent number of ATP molecules produced/consumed along the processes of interest \citep{cao_free-energy_2015, rodenfels_heat_2019, yang_physical_2021}. However, such purely thermodynamic approaches do not account for reaction kinetics and thus cannot quantify the rates of free energy transduction and dissipation. Significant efforts have been made in the direction of adding thermodynamic constraints in flux balance analysis of metabolic networks \citep{beard_energy_2002, niebel_upper_2019}. A few attempts have also been made to account for more complex kinetic effects such as enzyme saturation, leading to insights into the trade-offs between energy production and enzyme costs in glycolysis \citep{noor_note_2013, flamholz_glycolytic_2013, stettner_cost_2013}. Nevertheless, all these approaches rely on optimized nonequilibrium steady-states, which may not correspond to physiological conditions and cannot capture the energetic impact of time-dependent behaviors, such as the energy-rescuing effect of \ce{Ca^{2+}} oscillations quantified here. Our approach overcomes these limitations, based on the rigorous thermodynamic analysis of a curated dynamical model. Due to the modular structure of the model, our approach can be extended with additional pathways, such as glycolysis or one-carbon metabolism, with the aim to perform \textit{integrative modeling} of cell metabolism.

Finally, a key mechanism of the \ce{Ca^{2+}}-mitochondria crosstalk and the metabolic efficiency management is the dynamical switch between SERCA and mitochondrial uptakes. Alterations in \ce{Ca^{2+}} removal mechanisms due to mutations, generation of reactive oxygen species or remodeling of channel and pump expression are ubiquitous in pathological states such as mitochondrial \citep{visch_ca2-mobilizing_2006} and neurodegenerative diseases \citep{celsi_mitochondria_2009, filadi_mitochondrial_2020}, cancer \citep{monteith_calciumcancer_2017} or diabetes \citep{guerrero-hernandez_calcium_2014}, and therapeutic strategies targeting \ce{Ca^{2+}} homeostasis and signaling have started to emerge \citep{giorgi_mitochondrial_2012, dejos_two-way_2020}. Some of these changes can be captured by perturbations in the kinetic parameters of the \ce{Ca^{2+}} fluxes \citep{soman_inhibition_2017}, which would make the use of our approach in the context of disease quite straightforward. Overall, our methodology thus paves the way for a more systematic characterization of the dynamical energetic impact of metabolism regulation, which could improve the current understanding of pathway selection mechanisms in health and disease.


\section{Methods}
\label{sec:methods}

Model development, methodology and calculations are detailed in the next sections as follows: (1) Description of the kinetic model, (2) Concepts of biothermodynamics and (3) Nonequilibrium thermodynamics analysis. A systematic description of the nonequilibrium thermodynamics of CRNs can be found in \cite{rao_conservation_2018} and \cite{avanzini_nonequilibrium_2021}. This description can be applied to effective coarse-grained fluxes \citep{wachtel_thermodynamically_2018, avanzini_thermodynamics_2020} such as the ones we used to characterize the enzyme-catalysed processes of our system.
The chemical reactions incorporated in the model are listed in \TABLE{chemEqn} and the corresponding fluxes and forces are given in \TABLE{fluxes} and \TABLE{forces}, respectively. Lastly, the reference simulation parameters and a short description of their physical meaning can be found in \TABLE{param}.
Simulation algorithms are publicly available at \url{https://gitlab.lcsb.uni.lu/ICS-lcsb/net-ca-mito.git} as of the date of publication. Transformed Gibbs free energies of reaction were retrieved from eQuilibrator \citep{flamholz_equilibratorbiochemical_2012} and data were generated by running the simulation algorithms in Python (version 3.9.13).

\subsection{Kinetic model}
\subsubsection{Mitochondrial reactions}
Our model for mitochondrial metabolism is partly based on the pioneering work of Magnus and Keizer \citep{magnus_minimal_1997,magnus_model_1998-c,magnus_model_1998-m}. In particular, the synthesis of ATP by F1F0-ATPase and the oxidation of NADH into \ce{NAD^{+}} by the electron transport chain are both accompanied by a flux of protons across the mitochondrial membrane. While the reaction fluxes and the proton fluxes are described by slightly different expressions in the Magnus-Keizer model, we consider that the proton flux is a multiple of the reaction flux, and fix the proportionality coefficient according to the known stoichiometry (see \TABLE{chemEqn}), as done in previous models \citep{bertram_simplified_2006, wacquier_interplay_2016}. The Magnus-Keizer model accounts for mitochondrial carbon metabolism in a compact way, with effective fluxes for the TCA cycle and the associated generation of high energy electrons in the form of NADH. These fluxes have a constant component representing a basal flux, which is non-zero even in the absence of carbon input (glucose), and a contribution from glycolysis, which is assumed to be essentially captured by the flux through PDH \citep{magnus_model_1998-c}. Although these effective fluxes depend on $\left[\mathrm{Ca}^{2+}\right]_\mathrm{m}$ to account for the $\mathrm{Ca}^{2+}$-regulation of PDH, they do not reflect the activatory role of \ce{Ca^{2+}} on IDH and KGDH. We thus replace these effective contributions by incorporating the more detailed model of the TCA cycle from Dudycha \textit{et al.} \citep{dudycha_detailed_2000}. A reaction current is associated to each of the 8 enzymatic reaction of the TCA cycles, which are subjected to possible regulations by \ce{Ca^{2+}}, but also by \ce{Mg^{2+}}, ATP, NADH, and others. Additionally, Dudycha's model accounts for the aspartate aminotransferase reaction, which is not part of the TCA cycle \textit{per se}, but converts oxaloacetate and glutamate into $\alpha$-ketoglutarate and aspartate.
We neglect the aspartate aminotransferase reaction to focus exclusively on the TCA reactions.
Finally, we use acetyl-CoA (AcCoA) at the entry of the TCA cycle as the carbon input of the mitochondrial metabolism instead of glucose at the entry of glycolysis like in Magnus-Keizer model.
Note that these modifications are similar to the ones adopted by Cortassa \textit{et al.} in~\cite{cortassa_integrated_2003}.\\

\subsubsection{Cytosolic reactions and calcium signaling}
In~\cite{cortassa_integrated_2003}, $\left[\mathrm{Ca}^{2+}\right]_c$ is as a constant parameter and the ATP-consuming \ce{Ca^{2+}} exchanges between the endoplasmic reticulum (ER) and the cytosol are neglected. The resulting model can therefore not fully capture the crosstalk between mitochondrial ATP production and \ce{Ca^{2+}} signaling. To close this gap, our model explicitly describes the coupling between ATP hydrolysis and the uptake of Ca$^{2+}$ into the ER by SERCA pumps as done by Wacquier \textit{et al.} \citep{wacquier_interplay_2016}. More precisely, the SERCA flux is ATP-dependent and affects the cytosolic concentration of ATP. The \ce{Ca^{2+}} uptake by SERCAs is balanced by the efflux of \ce{Ca^{2+}} from the ER through \ce{IP3} receptors (IP$_3$Rs) and passive leak through the membrane of the ER. The flux through IP$_3$Rs is modeled by a function depending solely on the concentrations of \ce{IP3} and cytosolic \ce{Ca^{2+}}, as in~\cite{komin_multiscale_2015}.\\

\subsubsection{Fluxes and evolution equations}
Overall, we consider 32 species and 17 chemical reactions taking place in or between compartments corresponding to the mitochondrial matrix, the cytosol and the ER. The chemical equations describing these reactions can be found in \TABLE{chemEqn}. A current $J_\kappa$ (given in \TABLE{fluxes}) is associated to each chemical reaction $\kappa$ and is normalized with respect to the volume of the corresponding cell compartment (cytosol, ER or mitochondria). The concentration of the controlled species \{Pi\textsubscript{c}, Pi\textsubscript{m}, $\mathrm{Na}_\mathrm{c}^{+}$, $\mathrm{Na}_\mathrm{m}^{+}$, IP\textsubscript{3}, $\mathrm{H}_\mathrm{c}^{+}$, $\mathrm{H}_\mathrm{m}^{+}$, \ce{O2}, H\textsubscript{2}O\textsubscript{c}, H\textsubscript{2}O\textsubscript{m}, AcCoA, CoA, CO\textsubscript{2}, CoQH\textsubscript{2}, CoQ\} is fixed in time~(see also \FIG{fluxes} in the main text), meaning that the effect of chemical reactions is balanced by additional processes which are not described explicitly in the model. The rate equations for the other species are given in \EQ{firstODE}--\EQ{lastODE}. Stoichiometric and volumetric coefficients are included in \EQ{firstODE}--\EQ{lastODE} to guarantee mass balance across cell compartments. Specifically, $\alpha=V_\mathrm{ER}/V_\mathrm{c}$ and $\delta=V_\mathrm{m} / V_\mathrm{c}$, where $V_\mathrm{c}$, $V_\mathrm{ER}$ and $V_\mathrm{m}$ are the volumes of the cytosol, of the ER and of mitochondria, respectively. \ce{Ca^{2+}} buffering in these compartments is accounted for by the coefficients $f_\mathrm{c}$, $f_\mathrm{ER}$ and $f_\mathrm{m}$, which correspond to the fraction of free \ce{Ca^{2+}} in the compartment of interest. Finally, we mention that all fluxes are expressed in mM/s and concentration units are mM, except for \ce{Ca^{2+}} concentrations which are in $\mu$M, and coefficient $\gamma = 10^3$ $\mu$M/mM is therefore introduced to ensure consistency in units.

\begin{eqnarray}
    &\mathrm{d}_t \left[\mathrm{ADP}\right]_\mathrm{c} = -\delta J_\mathrm{ANT} + J_\mathrm{Hyd} + \frac{1}{2} J_\mathrm{SERCA} \label{eq:firstODE}\\
    &\mathrm{d}_t \left[\mathrm{ADP}\right]_\mathrm{m} = J_\mathrm{ANT} - J_\mathrm{F1} - J_\mathrm{SL}\\ 
    &\mathrm{d}_t \left[\mathrm{\alpha KG}\right]_\mathrm{m} = J_\mathrm{IDH} -J_\mathrm{KGDH} \\ 
    &\mathrm{d}_t \left[\mathrm{ATP}\right]_\mathrm{c} = \delta J_\mathrm{ANT} - J_\mathrm{Hyd} - \frac{1}{2} J_\mathrm{SERCA}\label{eq:ODEatpc}\\ 
    &\mathrm{d}_t \left[\mathrm{ATP}\right]_\mathrm{m} = -J_\mathrm{ANT} + J_\mathrm{F1} + J_\mathrm{SL} \\
    &\mathrm{d}_t \left[\mathrm{Ca^{2+}}\right]_\mathrm{c} = \frac{f_\mathrm{c}}{\gamma} \left(-J_\mathrm{SERCA} + J_\mathrm{ERout} + \delta \left(J_\mathrm{NCX} - J_\mathrm{UNI}\right)\right) \\
    &\mathrm{d}_t \left[\mathrm{Ca^{2+}}\right]_\mathrm{m} = \frac{f_\mathrm{m}}{\gamma} \left(J_\mathrm{UNI} - J_\mathrm{NCX}\right) \\
    &\mathrm{d}_t \left[\mathrm{Ca^{2+}}\right]_\mathrm{ER} = - \frac{f_\mathrm{ER}}{\gamma \alpha} \left(J_\mathrm{SERCA} - J_\mathrm{ERout}\right) \\
    &\mathrm{d}_t \left[\mathrm{CIT}\right]_\mathrm{m} = J_\mathrm{CS}- J_\mathrm{ACO} \\
    &\mathrm{d}_t \left[\mathrm{FUM}\right]_\mathrm{m} = J_\mathrm{SDH} - J_\mathrm{FH} \\
    &\mathrm{d}_t \left[\mathrm{ISOC}\right]_\mathrm{m} = J_\mathrm{ACO} - J_\mathrm{IDH} \\
    &\mathrm{d}_t \left[\mathrm{MAL}\right]_\mathrm{m} = J_\mathrm{FH}-J_\mathrm{MDH} \\
    &\mathrm{d}_t \left[\mathrm{NAD}\right]_\mathrm{m} = J_\mathrm{Ox} - J_\mathrm{IDH} - J_\mathrm{KGDH} - J_\mathrm{MDH} \\
    &\mathrm{d}_t \left[\mathrm{NADH}\right]_\mathrm{m} = -J_\mathrm{Ox} + J_\mathrm{IDH} + J_\mathrm{KGDH} +J_\mathrm{MDH} \\
    &\mathrm{d}_t \left[\mathrm{OAA}\right]_\mathrm{m} = J_\mathrm{MDH} - J_\mathrm{CS} \\
    &\mathrm{d}_t \Delta \Psi = \frac{1}{C_m}\left(10\, J_\mathrm{Ox} - 3\, J_\mathrm{F1} - J_\mathrm{ANT} - J_\mathrm{Hl} -J_\mathrm{NCX} - 2\, J_\mathrm{UNI}\right)\\
    &\mathrm{d}_t \left[\mathrm{SCoA}\right]_\mathrm{m} = J_\mathrm{KGDH} - J_\mathrm{SL} \\
    &\mathrm{d}_t \left[\mathrm{SUC}\right]_\mathrm{m} = J_\mathrm{SL} - J_\mathrm{SDH}. \label{eq:lastODE}
\end{eqnarray}


\subsection{Concepts of biothermodynamics}
\subsubsection{Definitions}
The entropy production rate (EPR) associated to a chemical reaction $\rho$ is given by
\begin{equation}
\sigma_{\rho}= - \frac{J_{\rho}\, \Delta_r G_{\rho}}{T} \geq 0, 
\label{eq_EPR}
\end{equation}
where $T$ is the absolute temperature while $J_\rho$ and $\Delta_r G_{\rho}$ are the current and Gibbs free energy of reaction $\rho$, respectively. In nonequilibrium thermodynamics, $-\Delta_r G_{\rho}$ is the \textit{force} driving the reaction while $J_\rho$ is the reaction \textit{flux} resulting from this force. The equilibrium state is characterized by zero forces and hence zero fluxes. 

The Gibbs free energy of reaction $\rho$ is defined by \citep{de_groot_non-equilibrium_1984}
\begin{equation}
\Delta_r G_{\rho}=\sum\limits_{i} \mathbb{S}_i^{\rho} \mu_{i},
\end{equation}
where $\mathbb{S}_i^\rho$ is the net stoichiometric coefficient of species $i$ in reaction $\rho$ and $\mu_i$ is the chemical potential of species $i$. Under the hypothesis of \textit{local equilibrium}, \textit{i.e}.,\ state variables such as temperature and pressure relax to equilibrium on a much faster timescale than chemical reactions, the expressions for the chemical potentials derived at equilibrium still hold locally out-of-equilibrium \citep{de_groot_non-equilibrium_1984}. The chemical potential $\mu_i$ is thus given by
\begin{equation}
\mu_i=\mu_i^\circ + R T \ln a_i,
\label{eq_chemical_potential}
\end{equation}
where $\mu_i^\circ$ and $a_i$ denote the standard chemical potential and the activity of species $i$, respectively. The activity accounts for the interactions between chemical species present in solution and is related to the concentration by the coefficient of activity $\gamma_i$, which depends on the ionic strength \citep{alberty_thermodynamics_2003}, $I$, such that $a_i = \gamma_i \frac{\left[i\right]}{c^\circ}$, where $\left[i\right]$ and $c^\circ$ are the concentration of species $i$ and standard concentration, respectively. In ideal solutions, $\gamma_i=1$. Standard conditions correspond to atmospheric pressure $p^\circ=1$ bar and molar concentrations $c^\circ=1\,\mathrm{M}$. 

Standard chemical potentials are directly related to the standard Gibbs free energy of reaction $\Delta_r G_\rho^\circ=\sum\limits_{i} \mathbb{S}_{i}^{\rho} \mu_{i}^\circ=- RT \ln K_{\rho}$, where $K_\rho = \prod\limits_{i} a_{i, eq}^{\mathbb{S}_{i}^{\rho}}$ is the equilibrium constant of reaction $\rho$. Hence, $\Delta_r G_{\rho}$ can then be rewritten as
\begin{equation}
\Delta_r G_\rho = \Delta_r G_\rho^\circ + R T \ln \prod\limits_{i} a_i^{\mathbb{S}_{i}^{\rho}} = R T \ln \prod\limits_{i} \left(\frac{a_i}{a_{i, eq}}\right)^{\mathbb{S}_{i}^{\rho}}.
\label{eq_deltaG}
\end{equation}
From a practical point of view, standard Gibbs free energies of reaction and activity coefficients are usually available in thermodynamic tables. As described in the next subsection, further adaptations can be done to describe more adequately the physiological environment in which biochemical reactions take place.

\subsubsection{Physiological conditions}
Cells are compartmentalized into specialized organelles whose composition can widely differ. For example, mitochondrial and cytosolic pH are 8 \citep{casey_sensors_2010} and 7.2 \citep{buckler_application_1990}, respectively. A plethora of buffering mechanisms regulate their internal concentration of metallic ions and pH, and thereby ensures the maintenance of homeostasis. Some chemical species can exist in different forms, that is, bound to metallic cations or at different levels of protonation (for example, ``ATP'' can be $\text{ATP}^{4-}$, $\text{HATP}^{3-}$, $\text{MgATP}^{2-}$, \textit{etc.}), and their relative abundance depends on the internal environment of the organelle.
For the sake of 
compactness, biochemical reactions are thus usually written in terms of \textit{pseudoisomers}, that is, without explicitly mentioning the state of the species and without detailing the consumption or production of protons/metallic ions by the reaction (e.g. $\text{ATP}+ \text{H}_2\text{O} \rightleftharpoons \text{ADP} + \text{P}_i$) \citep{alberty_thermodynamics_2003}.

To describe biochemical reactions from a thermodynamic point of view, their associated standard Gibbs free energy can be rescaled to match the equilibrium corresponding the physiological pH and metallic ions concentrations, but also incorporates the activity coefficients corresponding to a physiological ionic strength ($I=0.120\, \text{M}$ \citep{alberty_thermodynamics_2003, robinson_em_2006}). The resulting $\Delta_r G'^\circ_\kappa$ is subsequently used to calculate the transformed Gibbs free energy of reaction 
\begin{equation}
\Delta_r G'_\kappa = \Delta_r G'^\circ_\kappa + RT \ln \prod\limits_{j} \left[j\right]^{\mathbb{S}^{\kappa}_{j}},
\label{eq_transfGibbsEn}
\end{equation}
where $\left[j\right]$ is the concentration of pseudoisomer $j$ and $\mathbb{S}^{\kappa}_{j}$ is the net stoichiometric coefficient of pseudoisomer $j$ in reaction $\kappa$. Complementary approaches \citep{noor_integrated_2012, noor_consistent_2013} have led to the development of databases \citep{goldberg_thermodynamics_2004, flamholz_equilibratorbiochemical_2012} from which we retrieved $\Delta_r G'^\circ_\kappa$ for different pH, ionic strength and metallic ion concentrations. 

Electrogenic processes, such as $\text{Ca}^{2+}$ exchanges, the exchange of $\text{ATP}^{4-}$ and $\text{ADP}^{3-}$ \textit{via} the antiporter and the transfer of protons accompanying oxidative phosphorylation, constitute notable exceptions where the charges of species need to be explicitly accounted for. Indeed, electrostatic interactions affect the Gibbs free energy and in that case, the right-hand side of Eq.~\ref{eq_transfGibbsEn} must also comprise the term related to the electric potential in the compartment of interest. We thus distinguish the charged species $\{i\}$ from the pseudoisomers $\{j\}$. More precisely, $\Delta_r G'_\kappa = \sum\limits_{j} \mathbb{S}_j^{\kappa} \mu_{j} + \sum\limits_{i} \mathbb{S}_i^{\kappa} \bar{\mu}_{i}$ where $\bar{\mu}_{i}$ is the \textit{electrochemical potential}
\begin{equation}
\bar{\mu}_i=\mu_i^\circ + R T \ln a_i + z_i F V_{r\left(i\right)},
\label{eq_electrochemical_potential}
\end{equation}
where $z_i$ is the charge of species $i$ (for example, $z=+2$ for $\mathrm{Ca}^{2+}$), $F$ is the Faraday constant and $V_{r\left(i\right)}$ is the electric potential in the compartment $r$ where species $i$ is considered.
Overall, this leads to
\begin{equation}
    \Delta_r G'_\kappa = \Delta_r G'^\circ_\kappa + RT \ln \prod\limits_{j} \left[j\right]^{\mathbb{S}^{\kappa}_{j}} + RT \ln \prod\limits_{i} \left[i\right]^{\mathbb{S}^{\kappa}_{i}} + \sum\limits_{i}\mathbb{S}^{\kappa}_{i}z_i F V_{r\left(i\right)}.
\end{equation}
When a charged species is exchanged between two compartments, the $\Delta_r G'_\kappa$ associated to this transport process depends on the difference of potential between the two compartments. For example, if we consider the transport of $\mathrm{Ca}^{2+}$ from cytosol to mitochondria, $\Delta_r G'_\kappa = RT\ln  \frac{[\mathrm{Ca}^{2+}]_{\mathrm m}}{[\mathrm{Ca}^{2+}]_{\mathrm c}} + 2 F \left(V_m - V_c\right) = RT\ln  \frac{[\mathrm{Ca}^{2+}]_{\mathrm m}}{[\mathrm{Ca}^{2+}]_{\mathrm c}} - 2 F\Delta \psi$. 

Although pH (and hence proton concentrations) in mitochondria and cytosol are assumed to be constant due to strong buffering, proton transfer across mitochondrial membrane still affects the membrane potential (at least, at the local scale that is considered in the present model), which in turns affect $\Delta_r G'_\kappa$. The expressions for the $\Delta_r G'_\kappa$ of each process of the model can be found in \TABLE{forces}.


\subsection{\textbf{Nonequilibrium thermodynamic analysis}}
\subsubsection{Mitochondria as chemical engines}
Mitochondria can be considered as engines converting ADP\textsubscript{c} into ATP\textsubscript{c} \textit{via} 11 so-called \textit{internal reactions}
\begin{equation*}
    \kappa_i \in \{\text{ANT, F1, OX, CS, ACO, IDH, KGDH, SL, SDH, FH, MDH}\},
\end{equation*}
modeling mitochondrial metabolism, which are coupled to 6 \textit{external reactions} 
\begin{equation*}
    \kappa_e \in \{ \text{ERout, SERCA, NCX, UNI, HYD, Hl}\},
\end{equation*}
representing \ce{Ca^{2+}} signaling, cell activity and ionic homeostasis.
The chemical species involved in internal reactions are categorized into two groups, referred to as \textit{internal species}
\begin{equation*}
    {X} \in \{\mathrm{ATP}_\mathrm{m}, \mathrm{ADP}_\mathrm{m}, \mathrm{NADH}, \mathrm{NAD}, \mathrm{OAA}, \mathrm{CIT}, \mathrm{ISOC}, \alpha\mathrm{KG}, \mathrm{SCoA}, \mathrm{SUC}, \mathrm{FUM}, \mathrm{MAL}\},
\end{equation*}
and \textit{exchanged species}
\begin{equation*}
    {Y} \in \{ \mathrm{ATP}_\mathrm{c}, \mathrm{ADP}_\mathrm{c}, \mathrm{Pi}_\mathrm{m}, \mathrm{H}_\mathrm{c}^{+}, \mathrm{H}_\mathrm{m}^{+}, \mathrm{O}_2, \mathrm{H}_2\mathrm{O}_\mathrm{m}, \mathrm{AcCoA}, \mathrm{CoA}, \mathrm{CO}_2, \mathrm{CoQ}, \mathrm{CoQH}_2 \}.
\end{equation*}
The former is the set of species involved only in the internal reactions,
while the latter includes the controlled species and the species involved also in the external reactions.
Thus, the rate equations for internal and exchanged species can be respectively written as
\begin{subequations}
\begin{equation}
\mathrm{d}_t \left[\mathrm{X}\right] = \sum\limits_{\kappa_i} \mathbb{S}_{\kappa_i}^\mathrm{X} J_{\kappa_i}\,, \label{eq:dxdt}
\end{equation}
\begin{equation}
\frac{V_\mathrm{ref}^\mathrm{Y}}{V_\mathrm{m}} \mathrm{d}_t \left[\mathrm{Y}\right] = \sum\limits_{\kappa_i} \mathbb{S}_{\kappa_i}^\mathrm{Y} J_{\kappa_i} + I^\mathrm{Y} \label{eq:dydt},
\end{equation}
\label{eq:dynamics}
\end{subequations}
where $V_\mathrm{ref}^\mathrm{Y}$ is the volume of the compartment to which species Y belongs, and $I^\mathrm{Y}$ is the exchange current either accounting for the external reactions (named flux control in~\cite{avanzini_thermodynamics_2022}) or modeling additional processes that are responsible for the homeostasis of the controlled species (named concentration control in~\cite{avanzini_thermodynamics_2022}).
On the one hand $\mathrm{ATP}_\mathrm{c}$ and $\mathrm{ADP}_\mathrm{c}$ are involved in the external reactions SERCA and HYD and hence  $I^\mathrm{ATP_c} = -I^\mathrm{ADP_c} =\frac{1}{\delta}(-\frac{1}{2} J_\mathrm{SERCA} - J_\mathrm{HYD})$;
on the other hand, the other exchanged species (\textit{i.e.}, $\mathrm{Pi}_\mathrm{m}$, $\mathrm{H}^{+}_\mathrm{m}$, $\mathrm{H}^{+}_\mathrm{m}$,  $\mathrm{O}_2$, $\mathrm{H_2O}_\mathrm{m}$, $\mathrm{AcCoA}$, $\mathrm{CoA}$, $\mathrm{CO_2}$, $\mathrm{CoQH_2}$, $\mathrm{CoQ}$) are controlled species and hence $I^\mathrm{Y}=- \sum\limits_{\kappa_i} \mathbb{S}_{\kappa_i}^\mathrm{Y} J_{\kappa_i}$.

\subsubsection{Second law for mitochondrial metabolism\label{sec:2law}}
In general, the second law of thermodynamics for open CRNs can be written as~\citep{rao_conservation_2018, avanzini_nonequilibrium_2021} 
\begin{equation}
    T \sigma = - \mathrm{d}_t \mathcal{G} + \dot{w}_\mathrm{nc} + \dot{w}_\mathrm{driv},
    \label{eq:2law}
\end{equation}
where $\mathcal{G}$ is the (semi-grand) \textit{Gibbs free energy} of the system, while $\dot{w}_\mathrm{nc}$ and $\dot{w}_\mathrm{driv}$, respectively referred to as the \textit{nonconservative work rate} and the \textit{driving work rate}, are related to the energetic cost of maintaining CRNs out of equilibrium \textit{via} the exchange of species $\{\mathrm Y\}$.
Since $\mathcal{G}$ is a state function, its time derivative vanishes at steady state as well as when averaged over one period in the case of an oscillatory regime. \\

In the following, we use the topological analysis developed in~\cite{rao_conservation_2018, avanzini_nonequilibrium_2021} to derive the explicit expressions of the nonconservative work rate and the driving work rate for mitochondrial metabolism. \\

\textit{Remark:} The rate equations~(\EQ{firstODE}-\EQ{lastODE}) are coarse-grained, namely, each reactive process represents a sequence of out-of-equilibrium elementary reactions involving intermediate species whose dynamical behavior is not described.
Each of these elementary reactions might affect the energetics of the whole system.
Nevertheless, under the assumption of the existence of a time scale separation between the evolution of the species accounted by the dynamical model and the coarse-grained intermediate species, our thermodynamic analysis characterizes the correct energetics of the whole system as proven in~\cite{avanzini_thermodynamics_2020, avanzini_circuit_2023}.

\subsubsection{Conservation laws and emergent cycles}
For our model, the \textit{stoichiometric matrix} $\mathbb{S}$ encoding the net stoichiometric coefficients of internal species X and exchanged species Y in the internal reactions $\kappa_i$ reads
\renewcommand{\kbldelim}{(}
\renewcommand{\kbrdelim}{)}
\begin{equation}\label{eq:stoichioMatrix}
  \mathbb{S} = \text{ }\kbordermatrix{
    & \mathrm{ANT} & \mathrm{F1} & \mathrm{OX} & \mathrm{CS} & \mathrm{ACO} & \mathrm{IDH} & \mathrm{KGDH} & \mathrm{SL} & \mathrm{SDH} & \mathrm{FH} & \mathrm{MDH}\\
    \mathrm{ATP}_\mathrm{m} & -1 & 1 & 0 & 0 & 0 & 0 & 0 & 1 & 0 & 0 & 0 \\
    \mathrm{ADP}_\mathrm{m} & 1 & -1 & 0 & 0 & 0 & 0 & 0 & -1 & 0 & 0 & 0 \\
    \mathrm{NADH} & 0 & 0 & -1 & 0 & 0 & 1 & 1 & 0 & 0 & 0 & 1 \\
    \mathrm{NAD} & 0 & 0 & 1 & 0 & 0 & -1 & -1 & 0 & 0 & 0 & -1 \\
    \mathrm{OAA} & 0 & 0 & 0 & -1 & 0 & 0 & 0 & 0 & 0 & 0 & 1 \\
    \mathrm{CIT} & 0 & 0 & 0 & 1 & -1 & 0 & 0 & 0 & 0 & 0 & 0 \\
    \mathrm{ISOC} & 0 & 0 & 0 & 0 & 1 & -1 & 0 & 0 & 0 & 0 & 0 \\
    \alpha\mathrm{KG} & 0 & 0 & 0 & 0 & 0 & 1 & -1 & 0 & 0 & 0 & 0 \\
    \mathrm{SCoA} & 0 & 0 & 0 & 0 & 0 & 0 & 1 & -1 & 0 & 0 & 0 \\
    \mathrm{SUC} & 0 & 0 & 0 & 0 & 0 & 0 & 0 & 1 & -1 & 0 & 0 \\
    \mathrm{FUM} & 0 & 0 & 0 & 0 & 0 & 0 & 0 & 0 & 1 & -1 & 0 \\
    \mathrm{MAL} & 0 & 0 & 0 & 0 & 0 & 0 & 0 & 0 & 0 & 1 & -1 \\
    \hline
    \mathrm{ADP}_\mathrm{c} & -1 & 0 & 0 & 0 & 0 & 0 & 0 & 0 & 0 & 0 & 0 \\
    \mathrm{Pi}_\mathrm{m} & 0 & -1 & 0 & 0 & 0 & 0 & 0 & -1 & 0 & 0 & 0 \\
    \mathrm{H}^{+}_\mathrm{m} & 0 & 3 & -10 & 0 & 0 & 0 & 0 & 0 & 0 & 0 & 0 \\
    \mathrm{O}_2 & 0 & 0 & -\frac{1}{2} & 0 & 0 & 0 & 0 & 0 & 0 & 0 & 0 \\
    \mathrm{H_2O}_\mathrm{m} & 0 & 1 & 1 & -1 & 0 & 0 & 0 & 0 & 0 & -1 & 0 \\
    \mathrm{AcCoA} & 0 & 0 & 0 & -1 & 0 & 0 & 0 & 0 & 0 & 0 & 0 \\
    \mathrm{CoA} & 0 & 0 & 0 & 1 & 0 & 0 & -1 & 1 & 0 & 0 & 0 \\
    \mathrm{CO_2} & 0 & 0 & 0 & 0 & 0 & 1 & 1 & 0 & 0 & 0 & 0 \\
    \mathrm{CoQH_2} & 0 & 0 & 0 & 0 & 0 & 0 & 0 & 0 & 1 & 0 & 0 \\
    \mathrm{CoQ} & 0 & 0 & 0 & 0 & 0 & 0 & 0 & 0 & -1 & 0 & 0 \\
    \hline
    \mathrm{ATP}_\mathrm{c} & 1 & 0 & 0 & 0 & 0 & 0 & 0 & 0 & 0 & 0 & 0 \\
    \mathrm{H}^{+}_\mathrm{c} & 0 & -3 & 10 & 0 & 0 & 0 & 0 & 0 & 0 & 0 & 0
  }.
\end{equation}

The 13 (linearly-independent) left-null eigenvectors of $\mathbb{S}$, encoded as rows of the matrix

\begin{fullwidth}
    \renewcommand{\kbldelim}{(}
    \renewcommand{\kbrdelim}{)}
    \setlength\arraycolsep{1.2pt}
    \begin{equation}\label{eq:conserLaws}\small
        \mathbb{L} = \kbordermatrix{
        & \mathrm{ATP}_\mathrm{m} & \mathrm{ADP}_\mathrm{m} & \mathrm{NADH} &\mathrm{NAD} & \mathrm{OAA} & \mathrm{CIT} & \mathrm{ISOC} & \alpha\mathrm{KG} & \mathrm{SCoA} & \mathrm{SUC} & \mathrm{FUM} & \mathrm{MAL} &  & \mathrm{ADP}_\mathrm{c} & \mathrm{Pi}_\mathrm{m} & \mathrm{H}^{+}_\mathrm{m} & \mathrm{O}_2 & \mathrm{H_2O}_\mathrm{m} & \mathrm{AcCoA} & \mathrm{CoA} & \mathrm{CO_2} & \mathrm{CoQH_2} & \mathrm{CoQ} & & \mathrm{ATP}_\mathrm{c} & \mathrm{H}^{+}_\mathrm{c}\\
        \mathrm{H}^{+}_\mathrm{m}    & 0 & 0 & 0 & 0 & 0 & 0 & 0 & 0 & 0 & 0 & 0 & 0 & \vrule & 0 & 0 & 1 & 0 & 0 & 0 & 0 & 0 & 0 & 0 & \vrule & 0 & 1 \\
        \mathrm{ADP}_\mathrm{c}    & 0 & 0 & 0 & 0 & 0 & 0 & 0 & 0 & 0 & 0 & 0 & 0 & \vrule & 1 & 0 & 0 & 0 & 0 & 0 & 0 & 0 & 0 & 0 & \vrule & 1 & 0 \\
        \mathrm{CoQ}    & 3 & 0 & 10 & 0 & -10 & -10 & -10 & -20 & -30 & -33 & 0 & 0 & \vrule & -3 & 0 & -1 & 0 & 0 & 0 & 0 & 0 & 0 & 33 & \vrule & 0 & 0 \\
        \mathrm{CoQH_2}    & -3 & 0 & -10 & 0 & 10 & 10 & 10 & 20 & 30 & 33 & 0 & 0 & \vrule & 3 & 0 & 1 & 0 & 0 & 0 & 0 & 0 & 33 & 0 & \vrule & 0 & 0 \\
        \mathrm{CO_2}    & -6 & 0 & -20 & 0 & 20 & 20 & 20 & 7 & -6 & 0 & 0 & 0 & \vrule & 6 & 0 & 2 & 0 & 0 & 0 & 0 & 33 & 0 & 0 & \vrule & 0 & 0 \\
        \mathrm{CoA}    & -3 & 0 & -10 & 0 & 10 & -23 & -23 & -13 & 30 & 0 & 0 & 0 & \vrule & 3 & 0 & 1 & 0 & 0 & 0 & 33 & 0 & 0 & 0 & \vrule & 0 & 0 \\
        \mathrm{AcCoA}    & 3 & 0 & 10 & 0 & -10 & 23 & 23 & 13 & 3 & 0 & 0 & 0 & \vrule & -3 & 0 & -1 & 0 & 0 & 33 & 0 & 0 & 0 & 0 & \vrule & 0 & 0 \\
        \mathrm{H_2O}    & -1 & 0 & 1 & 0 & -1 & 0 & 0 & -1 & -2 & -1 & -1 & 0 & \vrule & 1 & 0 & 0 & 0 & 1 & 0 & 0 & 0 & 0 & 0 & \vrule & 0 & 0 \\
        \mathrm{O_2}    & 3 & 0 & -1 & 0 & 1 & 1 & 1 & 2 & 3 & 0 & 0 & 0 & \vrule & -3 & 0 & -1 & 22 & 0 & 0 & 0 & 0 & 0 & 0 & \vrule & 0 & 0 \\
        \mathrm{Pi}_\mathrm{m}    & 1 & 0 & 0 & 0 & 0 & 0 & 0 & 0 & 0 & 0 & 0 & 0 & \vrule & -1 & 1 & 0 & 0 & 0 & 0 & 0 & 0 & 0 & 0 & \vrule & 0 & 0 \\
        \hline
        \mathrm{OOA}    & 0 & 0 & 0 & 0 & 1 & 1 & 1 & 1 & 1 & 1 & 1 & 1 & \vrule & 0 & 0 & 0 & 0 & 0 & 0 & 0 & 0 & 0 & 0 & \vrule & 0 & 0 \\
        \mathrm{NADH}    & 0 & 0 & 1 & 1 & 0 & 0 & 0 & 0 & 0 & 0 & 0 & 0 & \vrule & 0 & 0 & 0 & 0 & 0 & 0 & 0 & 0 & 0 & 0 & \vrule & 0 & 0 \\
        \mathrm{ADP}_\mathrm{m}    & 1 & 1 & 0 & 0 & 0 & 0 & 0 & 0 & 0 & 0 & 0 & 0 & \vrule & 0 & 0 & 0 & 0 & 0 & 0 & 0 & 0 & 0 & 0 & \vrule & 0 & 0
        }\,,
\end{equation}
\end{fullwidth}

which therefore satisfies $\mathbb{L} \mathbb{S}=0$, define the conservation laws. 
Indeed, for the every row $\lambda$  of $\mathbb{L}$ (labeled using chemical symbols in \EQ{conserLaws} for reasons that will be explained in subsection~\nameref{sub:pot_for}), the quantity $L^\lambda = \sum_{\mathrm X}\mathbb{L}^\lambda_{\mathrm X} [\mathrm X]+ \sum_{\mathrm Y}\mathbb{L}^\lambda_{\mathrm Y} [\mathrm Y]$ would be a conserved quantity if mitochondria were closed systems, namely, if ${I}^\mathrm{Y}=0$~$\forall\, \mathrm Y$.
When ${I}^\mathrm{Y}\neq0$, only 3 out of the 13 conservation laws corresponding to the last three rows of $\mathbb{L}$ in \EQ{conserLaws} involve exclusively internal species and their corresponding quantities $L^\lambda =\sum_{\mathrm X}\mathbb{L}^\lambda_{\mathrm X} [\mathrm X]$ are still conserved. 
These conservation laws are said to be \textit{unbroken}. 
The other conservation laws correspond to quantities $L^\lambda  = \sum_{\mathrm X}\mathbb{L}^\lambda_{\mathrm X} [\mathrm X]+ \sum_{\mathrm Y}\mathbb{L}^\lambda_{\mathrm Y} [\mathrm Y]$ that are not conserved anymore and are, therefore, named \textit{broken} conservation laws.

The 2 (linearly-independent) right-null eigenvectors of $\mathbb{S}^{\mathrm X}$ (\textit{i.e.}, the (sub)\-stoichiometric matrix for the internal species), 
\renewcommand{\kbldelim}{(}
\renewcommand{\kbrdelim}{)}
\[
  \boldsymbol{c}_\mathrm{r1} = \kbordermatrix{
    & \mathrm{ANT} & \mathrm{F1} & \mathrm{OX} & \mathrm{CS} & \mathrm{ACO} & \mathrm{IDH} & \mathrm{KGDH} & \mathrm{SL} & \mathrm{SDH} & \mathrm{FH} & \mathrm{MDH}\\
    & 1 & \frac{10}{11} & \frac{3}{11} & \frac{1}{11} & \frac{1}{11} & \frac{1}{11} & \frac{1}{11} & \frac{1}{11} & \frac{1}{11} & \frac{1}{11} & \frac{1}{11}
  }\,,
\]
and
\[
  \boldsymbol{c}_\mathrm{r2} = \kbordermatrix{
    & \mathrm{ANT} & \mathrm{F1} & \mathrm{OX} & \mathrm{CS} & \mathrm{ACO} & \mathrm{IDH} & \mathrm{KGDH} & \mathrm{SL} & \mathrm{SDH} & \mathrm{FH} & \mathrm{MDH}\\
    & 0 & -\frac{1}{33} & \frac{1}{11} & \frac{1}{33} & \frac{1}{33} & \frac{1}{33} & \frac{1}{33} & \frac{1}{33} & \frac{1}{33} & \frac{1}{33} & \frac{1}{33}
  }\,,
\]
are named \textit{emergent cycles}.
They define sequences of reactions that overall leave the abundances of the internal species unchanged, since $\mathbb{S}^\mathrm{X}\boldsymbol{c}_\mathrm{r1}=0$ and $\mathbb{S}^\mathrm{X}\boldsymbol{c}_\mathrm{r2}=0$ by definition,
while interconverting the exchanged species as they undergo the effective reactions
\begin{align}
     \mathrm{ADP}_\mathrm{c} + \mathrm{Pi}_\mathrm{m} + \frac{3}{22}\, \mathrm{O}_2 + \frac{1}{11}\, \mathrm{AcCoA} + \frac{1}{11}\, \mathrm{CoQ} & \xrightleftharpoons[]{\mathbf{r1}} \mathrm{ATP}_\mathrm{c} + \mathrm{H}_2 \mathrm{O}_\mathrm{m} + \frac{1}{11}\, \mathrm{CoA} + \frac{2}{11}\, \mathrm{CO}_2 + \frac{1}{11}\, \mathrm{CoQH}_2 \label{eq:emATPc}\,,\\
    \mathrm{H}^{+}_\mathrm{m} +  \frac{1}{22}\, \mathrm{O}_2 + \frac{1}{33}\, \mathrm{AcCoA} + \frac{1}{33}\, \mathrm{CoQ} & \xrightleftharpoons[]{\mathbf{r2}} \mathrm{H}^{+}_\mathrm{c} + \frac{1}{33}\, \mathrm{CoA} + \frac{2}{33}\, \mathrm{CO}_2 + \frac{1}{33}\, \mathrm{CoQH}_2 \label{eq:emHc}\,,
\end{align}
whose stoichiometry is defined by $\mathbb{S}^\mathrm{Y}\boldsymbol{c}_\mathrm{r1}$ and $\mathbb{S}^\mathrm{Y}\boldsymbol{c}_\mathrm{r2}$, respectively, with $\mathbb{S}^{\mathrm Y}$ the (sub)stoichiometric matrix for the exchanged species.

\subsubsection{Potential and force species\label{sub:pot_for}}
The exchanged species $\{\mathrm Y\}$ can be classified as either \textit{potential} species $\{\mathrm Y_p\}$ or as \textit{force} species $\{\mathrm Y_f\}$.
The potential species are the largest subset of exchanged species such that the submatrix $\mathbb{L}^{b}_{\mathrm{Y}_p}$ of $\mathbb{L}$ for the broken conservation laws and the potential species can be inverted, \textit{i.e.}, $(\mathbb{L}^{b}_{\mathrm{Y}_p})^{-1}$ exists.
The force species are the remaining species $\{\mathrm Y_f\}=\{\mathrm Y\}\setminus\{\mathrm Y_p\}$.
As discussed in~\cite{rao_conservation_2018, avanzini_nonequilibrium_2021}, this partitioning is not unique, but different choices do not change the conclusions of the thermodynamic analysis.

In our model, we choose
\begin{equation*}
   \mathrm {Y}_p \in \{\mathrm{ADP}_\mathrm{c}, \mathrm{Pi}_\mathrm{m}, \mathrm{H}_\mathrm{m}^{+}, \mathrm{O}_2, \mathrm{H}_2\mathrm{O}_\mathrm{m}, \mathrm{AcCoA}, \mathrm{CoA}, \mathrm{CO}_2, \mathrm{CoQ}, \mathrm{CoQH}_2 \}
\end{equation*}
and 
\begin{equation*}
   \mathrm {Y}_f \in \{\mathrm{ATP}_\mathrm{c},\,\mathrm{H}_\mathrm{c}^+\}.
\end{equation*}
In Eq.~\ref{eq:stoichioMatrix}, the horizontal lines split $\mathbb{S}$ into 
\begin{equation}
\mathbb{S} =
    \begin{pmatrix}
    \mathbb{S}^{\mathrm{X}}\\
    \hline
    \mathbb{S}^{\mathrm{Y}_p} \\
    \hline 
     \mathbb{S}^{\mathrm{Y}_f} \\
    \end{pmatrix}
\end{equation}
In Eq.~\ref{eq:conserLaws} the horizontal and vertical lines split $\mathbb{L}$ into
\begin{equation}
\mathbb{L} =
    \begin{pmatrix}
    \mathbb{L}^{b}_{\mathrm{X}} & \vrule & \mathbb{L}^{b}_{\mathrm{Y}_p} & \vrule & \mathbb{L}^{b}_{\mathrm{Y}_f} \\
    \hline
    \mathbb{L}^{u}_{\mathrm{X}} & \vrule & \mathbb{L}^{u}_{\mathrm{Y}_p} & \vrule & \mathbb{L}^{u}_{\mathrm{Y}_f} \\
    \end{pmatrix}
\end{equation}
with $u$ and $b$ for unbroken and broken conservation laws, respectively. 

Note that the existence of $(\mathbb{L}^{b}_{\mathrm{Y}_p})^{-1}$ defines a representation of the broken conservation laws given by $(\mathbb{L}^{b}_{\mathrm{Y}_p})^{-1}\mathbb{L}^{b}$ where every broken conservation law involves only one potential species.
This physically means that each quantity $L^\lambda$ corresponding to a broken conservation law stops being conserved once a specific potential species is exchanged.
Furthermore, no new quantities $L^\lambda $ stop being conserved when the force species are exchanged. 
For this reason, each conservation law $\lambda$ in \EQ{conserLaws} can always be labeled using the chemical symbol of the potential species that, once exchanged, would make $L^\lambda$ a nonconserved quantity. 

\subsubsection{Nonconservative work}
The general expression of the nonconservative work rate is given by 
\begin{equation}
\dot{w}_\mathrm{nc} = \boldsymbol{\mathcal{F}}_{\mathrm{Y}_f}\cdot \boldsymbol{I}^{\mathrm{Y}_f}
\end{equation}
where
\begin{equation}
\boldsymbol{\mathcal{F}}_{\mathrm{Y}_f} = \big(\boldsymbol{\mu}_{\mathrm{Y}_f}\cdot \mathbb{I} - \boldsymbol{\mu}_{\mathrm{Y}_p}\cdot \left(\mathbb{L}^{b}_{\mathrm{Y}_p}\right)^{-1} \mathbb{L}^{b}_{\mathrm{Y}_f} \big)^\top
\end{equation}
is the vector of the nonconservative forces, $\boldsymbol{I}^{\mathrm{Y}_f}$ is the vector collecting the exchange currents of the force species, and $\boldsymbol{\mu}_{\mathrm{Y}_f}$ (resp. $\boldsymbol{\mu}_{\mathrm{Y}_p}$) is the vector of the chemical/electrochemical potential of the force (resp. potential) species. 

In our model, there are only two force species, namely, $\mathrm{ATP}_\mathrm{c}$ and $\mathrm{H}_\mathrm{c}^+$.
The two corresponding nonconservative forces are given by
\begin{align}
    \mathcal{F}_\mathrm{ATPc} &= -\frac{\text{$\mu_\mathrm{AcCoa}$}}{11}-\text{$\mu_\mathrm{ADP_c}$}+\text{$\mu_\mathrm{ATP_c}$}+\frac{2\, \text{$\mu_\mathrm{CO_2}$}}{11}+\frac{\text{$\mu_\mathrm{CoA}$}}{11}-\frac{\text{$\mu_\mathrm{CoQ}$}}{11}+\frac{\text{$\mu_\mathrm{CoQH_2}$}}{11}+\text{$\mu_\mathrm{H_2O_m}$}-\frac{3\, \text{$\mu_\mathrm{O_2}$}}{22}-\text{$\mu_\mathrm{Pim}$}
\label{eq:ncforce1a}
\end{align}
and
\begin{align}
    \mathcal{F}_\mathrm{H_c^+} &= -\frac{\text{$\mu_\mathrm{AcCoa}$}}{33}+\frac{2\, \text{$\mu_\mathrm{CO_2}$}}{33}+\frac{\text{$\mu_\mathrm{CoA}$}}{33}-\frac{\text{$\mu_\mathrm{CoQ}$}}{33}+\frac{\text{$\mu_\mathrm{CoQH_2}$}}{33}+\text{$\mu_\mathrm{H_c}$}-\text{$\mu_\mathrm{H_m}$}-\frac{\text{$\mu_\mathrm{O_2}$}}{22}
\label{eq:ncforce2a}
\end{align}
while the exchange currents are
\begin{equation}
    I^\mathrm{ATP_c} = \frac{1}{\delta} \left(- J_\mathrm{Hyd} - \frac{1}{2} J_\mathrm{SERCA}\right)\,,\label{eq:exc_atp}
\end{equation}
and
\begin{equation}
    I^\mathrm{H_c^+} = 3\, J_\mathrm{F1} - 10\, J_\mathrm{OX}\,.\label{eq:exc_h}
\end{equation}
\EQ{exc_atp} is obtained by writing the rate equation for $\mathrm{ATP_c}$ (\EQ{ODEatpc}) according to \EQ{dydt}: $\frac{1}{\delta}  \mathrm{d}_t \left[\mathrm{ATP}\right]_\mathrm{c} = J_\mathrm{ANT} + I^\mathrm{ATP_c}$.
\EQ{exc_h} is obtained by recognising that $\mathrm{H_c^+}$ is a controlled species, \textit{i.e.}, $\mathrm{d}_t \left[\mathrm{H^+}\right]_\mathrm{c} = 0$, and using again \EQ{dydt}, we can write $\frac{1}{\delta} \mathrm{d}_t \left[\mathrm{H^+}\right]_\mathrm{c} = 0 = 10\, J_\mathrm{OX} - 3\, J_\mathrm{F1} + I^\mathrm{H_c^+}$. 

Notice that the nonconservative forces in \EQ{ncforce1a} and~\EQ{ncforce2a} correspond exactly to $\Delta_r G'_\mathrm{r1}$ and $\Delta_r G'_\mathrm{r2}$ of the effective reactions in \EQ{emATPc} and~\EQ{emHc}.
They can therefore be rewritten as
\begin{align}
    \mathcal{F}_\mathrm{ATPc} 
    &= \Delta_r G'_\mathrm{ANT} + \frac{10\, \Delta_r G'_\mathrm{F1}}{11} + \frac{3\,\Delta_r G'_\mathrm{OX}}{11} + \frac{\Delta_r G'_\mathrm{TCA}}{11}= \Delta_r G'_\mathrm{r1}
\label{eq:ncforce1b}
\end{align}
and
\begin{align}
    \mathcal{F}_\mathrm{H_c^+} 
   &= -\frac{\Delta_r G'_\mathrm{F1}}{33} + \frac{\Delta_r G'_\mathrm{OX}}{11} + \frac{\Delta_r G'_\mathrm{TCA}}{33} = \Delta_r G'_\mathrm{r2}
\label{eq:ncforce2b}
\end{align}
where $\Delta_r G'_\mathrm{TCA} := \Delta_r G'_\mathrm{CS} + \Delta_r G'_\mathrm{ACO} + \Delta_r G'_\mathrm{IDH} + \Delta_r G'_\mathrm{KGDH} + \Delta_r G'_\mathrm{SL}+ \Delta_r G'_\mathrm{SDH} + \Delta_r G'_\mathrm{FH} + \Delta_r G'_\mathrm{MDH}$.
We numerically compute the nonconservative forces using the latter expressions.
The expression of the nonconservative work rate for our model then becomes
\begin{equation}
    \dot{w}_\mathrm{nc} =\Delta_r G'_\mathrm{r1} I^\mathrm{ATP_c} + \Delta_r G'_\mathrm{r2} I^\mathrm{H_c^+}\,.
    \label{eq:spec_ncw}
\end{equation}

\subsubsection{Driving work}
The driving work rate is in general given by the sum of two contributions,
\begin{equation}
\dot{w}_\mathrm{driv} = \dot{w}_\mathrm{driv}^\mathrm{ch} +  \dot{w}_\mathrm{driv}^\mathrm{in}\,,
\end{equation}
namely, the \textit{chemical driving work rate} 
\begin{equation}
    \dot{w}_\mathrm{driv}^\mathrm{ch} = - \left(\mathrm{d}_t \boldsymbol{\mu}_{\mathrm{Y}_p}\right) \cdot \left(\mathbb{L}^{b}_{\mathrm{Y}_p}\right)^{-1} {\mathbb{L}^{b} \left[\mathbf{Z}\right]},
\end{equation}
and the \textit{interaction driving work rate}
\begin{equation}
    \dot{w}_\mathrm{driv}^\mathrm{in} = \nabla_{\left[\boldsymbol{e}\right]} G^\mathrm{in}\left(\left[\mathbf{Z}\right], \left[\boldsymbol{e}\right]\right) \cdot \mathrm{d}_t \left[\boldsymbol{e}\right],
\end{equation}
Here, $[\mathbf{Z}] =([\mathbf{X}], [\mathbf{Y}]) $ is a vector collecting the concentrations of both internal and exchanged species, 
$\left[\boldsymbol{e}\right]$ is a vector collecting the concentrations of other interacting species (\textit{i.e.}, $\mathrm{Na}_\mathrm{m}^+$,  $\mathrm{Ca}_\mathrm{m}^{2+}$) that are not interconverted by the internal reactions and that do not appear in $\mathbb{S}$, 
$\nabla_{\left[\boldsymbol{e}\right]}$ is the gradient with respect to $\left[\boldsymbol{e}\right]$ 
and $G^\mathrm{in}\left(\left[\mathbf{Z}\right], \left[\boldsymbol{e}\right]\right)$ is the interaction Gibbs free energy, whose exact expression depends on the model used to describe interactions \citep{avanzini_nonequilibrium_2021}.

In our model, we compute the driving work rate over one period $t_p$ in the oscillatory regime (as it vanishes at steady state).
The specific expression of the average chemical driving work rate is given by
\begin{align}
    \frac{1}{t_p} \int_0^{t_p} \mathrm{d}t\, \dot{w}_\mathrm{driv}^\mathrm{ch} =& \frac{1}{t_p} \int_0^{t_p} \mathrm{d}t\, \left(- \mathrm{d}_t \mu_\mathrm{ADP_c} L^\mathrm{ADP_c} - \mathrm{d}_t \mu_\mathrm{Pi_m} L^\mathrm{Pi_m} - \mathrm{d}_t \mu_\mathrm{H_m^+} L^\mathrm{H_m^+} - \mathrm{d}_t \mu_\mathrm{O_2} L^\mathrm{O_2} /22- \mathrm{d}_t \mu_\mathrm{H_2O} L^\mathrm{H_2O} \right. \nonumber\\
    & \left.- \mathrm{d}_t \mu_\mathrm{AcCoA} L^\mathrm{AcCoA}/33 - \mathrm{d}_t \mu_\mathrm{CoA} L^\mathrm{CoA}/33 - \mathrm{d}_t \mu_\mathrm{CO_2} L^\mathrm{CO_2}/33 - \mathrm{d}_t \mu_\mathrm{CoQH_2} L^\mathrm{CoQH_2} /33- \mathrm{d}_t \mu_\mathrm{CoQ} L^\mathrm{CoQ}/33 \right).
\end{align}
Notice that the terms corresponding to uncharged controlled species (\textit{i.e.}, Pi\textsubscript{m}, O\textsubscript{2}, H\textsubscript{2}O\textsubscript{m}, AcCoA, CoA, CO\textsubscript{2}, CoQH\textsubscript{2}, CoQ) vanish since their chemical potential is constant over time.
For the charged controlled species $\mathrm{H}_\mathrm{m}^{+}$, the quantity $L^\mathrm{{H_m^+}}$ is still conserved since $\left[\mathrm{H}^{+}\right]_\mathrm{m}$ and $\left[\mathrm{H}^{+}\right]_\mathrm{c}$ are constant. 
Hence, $\frac{1}{t_p} \int_0^{t_p} \mathrm{d}t\, \left(\mathrm{d}_t \mu_\mathrm{H_m^+} L^\mathrm{H_m^+}\right) = \frac{L^\mathrm{H_m^+}}{t_p} \int_0^{t_p} \mathrm{d}t\, \left(\mathrm{d}_t \mu_\mathrm{H_m^+}\right)=0$ since the electrochemical potential is a state function. 
Similarly, $L^\mathrm{ADP_c} = \left[\mathrm{ADP}\right]_\mathrm{c}+ \left[\mathrm{ATP}\right]_\mathrm{c}$ is still conserved in the open system implying $\frac{1}{t_p} \int_0^{t_p} \mathrm{d}t\, \left(- \mathrm{d}_t \mu_\mathrm{ADP_c} L^\mathrm{ADP_c}\right)=\frac{L^\mathrm{ADP_c}}{t_p} \int_0^{t_p} \mathrm{d}t\, \left(- \mathrm{d}_t \mu_\mathrm{ADP_c}\right)=0$. 
In conclusion, the chemical driving work rate over one period vanishes:
\begin{equation}
 \frac{1}{t_p} \int_0^{t_p} \mathrm{d}t\, \dot{w}_\mathrm{driv}^\mathrm{ch} =0\,.
\end{equation}

We cannot determine the explicit expression of the interacting driving work rate since our model does not provide the interaction Gibbs free energy $G^\mathrm{in}\left(\left[\mathbf{Z}\right], \left[\boldsymbol{e}\right]\right)$.
Thus, we compute the driving work over one period by calculating the total entropy production rate as the sum of the individual EPR of each internal reaction $\boldsymbol{\kappa_i} $= \{ANT, F1, OX, CS, ACO, IDH, KGDH, SL, SDH, FH, MDH\},
\begin{equation}
T\sigma = - \sum\limits_{\boldsymbol{\kappa_i}} \Delta_r G'_{\kappa_i} J_{\kappa_i}
\end{equation}
from which we subtract the nonconservative work over one period:
\begin{equation}
     \frac{1}{t_p} \int_0^{t_p} \mathrm{d}t\, \dot{w}_\mathrm{driv} = \frac{1}{t_p} \int_0^{t_p} \mathrm{d}t\, \left(T\sigma - \dot{w}_\mathrm{nc}\right)\,.
\end{equation}

\subsubsection{Thermodynamic efficiency}
In subsection~\nameref{sec:2law}, we determined the specific expressions of work rates in the second law of thermodynamics~(\EQ{2law}), accounting for the free energy exchanges between mitochondria and their surroundings that balance dissipation and maintain the mitochondrial metabolism out of equilibrium.
The thermodynamic efficiency of mitochondrial metabolism is defined by identifying which of these terms play the role of free energy input and output.
To do so, we further split the nonconservative work rate~(\EQ{spec_ncw}) by recognizing that 
the effective reaction~(\EQ{emATPc}) is given by the sum of 2 (mass balanced) reactions
\begin{align}
    \mathrm{ADP}_\mathrm{c} + \mathrm{Pi}_\mathrm{m} & \xrightleftharpoons[]{\mathrm{\mathbf{r1_{out}}}} \mathrm{ATP}_\mathrm{c} + \mathrm{H}_2 \mathrm{O}_\mathrm{m} \label{eq:chemOutput}\,,\\
    \frac{3}{22}\, \mathrm{O}_2 + \frac{1}{11}\, \mathrm{AcCoA} + \frac{1}{11}\, \mathrm{CoQ} & \xrightleftharpoons[]{\mathrm{\mathbf{r1_{in}}}} \frac{1}{11}\, \mathrm{CoA} + \frac{2}{11}\, \mathrm{CO}_2 + \frac{1}{11}\, \mathrm{CoQH}_2 \label{eq:chemInputTCA}\,.
\end{align}
The effective reaction~\EQ{chemOutput} corresponds to the net production (output) of free energy (ATP\textsubscript{c}) by mitochondrial metabolism,
while the effective reactions~\EQ{chemInputTCA} and~\EQ{emHc} represent a free energy input generated by the chemical potential gradient between oxygen, coenzymes, protons and carbon substrates.
This allows us to write the nonconservative work rate~(\EQ{spec_ncw}) as
\begin{equation}
    \dot{w}_\mathrm{nc} = \Delta_r G'_\mathrm{r1_{in}} I^\mathrm{ATP_c}  + \Delta_r G'_\mathrm{r1_{out}} I^\mathrm{ATP_c} + \Delta_r G'_\mathrm{r2} I^\mathrm{H_c^+}\,.
    \label{eq:spec_final_ncw}
\end{equation}
The thermodynamic efficiency at steady state is thus given by
\begin{equation}
    \eta = - \frac{\Delta_r G'_\mathrm{r1_{out}} I^\mathrm{ATP_c}}{\Delta_r G'_\mathrm{r1_{in}}I^\mathrm{ATP_c}+\Delta_r G'_\mathrm{r2}I^\mathrm{H^+_c}}\,.
\end{equation}
In the oscillatory regime we also have to account for the free energy provided by the driving work and, therefore, the thermodynamic efficiency averaged over a period is
\begin{equation}
    \eta_{t_p} = - \frac{\int_0^{t_p} dt\, \Delta_r G'_\mathrm{r1_{out}} I^\mathrm{ATP_c}}{\int_0^{t_p} dt\, \left(\Delta_r G'_\mathrm{r1_{in}}I^\mathrm{ATP_c}+\Delta_r G'_\mathrm{r2}I^\mathrm{H^+_c}+ \dot{w}_\mathrm{driv}\right)}\,.
\end{equation}
In both cases, the thermodynamic efficiency quantify the amount of energy released by the synthesis of ATP normalized by the amount of energy injected in the mitochondria.
In the main text, we use the notation $\bar{\eta}$ to refer to the average efficiency at steady-state or in the oscillatory regime.

\begin{table}
\renewcommand{\arraystretch}{2.0}
\centering
\caption{Chemical reactions incorporated into the kinetic model. Controlled species are in gray. Subscripts c, ER and m refer to the cytosol, the endoplasmic reticulum and the mitochondria, respectively.}
\label{tab:chemEqn}
\begin{tabular}{lc}
\midrule
\multicolumn{2}{l}{\textbf{Calcium exchanges and cytosolic ATP dynamics}} \\
\midrule
ERout & $\text{Ca}^{2+}_\text{ER} \xrightleftharpoons[]{{\color{gray}\text{IP}_3}} \text{Ca}^{2+}_\text{c}$ \\
SERCA & $2\, \text{Ca}^{2+}_\text{c} + \text{ATP}_\mathrm{c} + {\color{gray}\text{H}_2 \text{O}_\text{c}} \rightleftharpoons 2\, \text{Ca}^{2+}_\text{ER} + \text{ADP}_\text{c} + {\color{gray} \text{Pi}_\text{c}}$\\
NCX & $\text{Ca}^{2+}_\text{m} + 3\, {\color{gray}\text{Na}^{+}_\text{c}} \rightleftharpoons \text{Ca}^{2+}_\text{c} + 3\, {\color{gray}\text{Na}^{+}_\text{m}}$\\
UNI & $\text{Ca}^{2+}_\text{c} \xrightleftharpoons[]{} \text{Ca}^{2+}_\text{m}$ \\
Hyd & $\text{ATP}_\mathrm{c} + {\color{gray}\text{H}_2 \text{O}_\text{c}} \rightleftharpoons \text{ADP}_\text{c} + {\color{gray}\text{Pi}_\text{c}}$\\
\midrule
\multicolumn{2}{l}{\textbf{Electron transport chain and oxidative phosphorylation}} \\
\midrule
OX & $\text{NADH} + 10\, {\color{gray}\text{H}^{+}_\text{m}} + \frac{1}{2} {\color{gray}\text{O}_2} \rightleftharpoons \text{NAD} + 10\, {\color{gray}\text{H}^{+}_\text{c}} + {\color{gray}\text{H}_2 \text{O}_\text{m}}$\\
F1 & $\text{ADP}_\text{m} + {\color{gray}\text{Pi}_\text{m}} + 3\, {\color{gray}\text{H}^{+}_\text{c}} \rightleftharpoons \text{ATP}_\text{m} + {\color{gray}\text{H}_2\text{O}_\text{m}} + 3\, {\color{gray}\text{H}^{+}_\text{m}}$\\
\midrule
\multicolumn{2}{l}{\textbf{TCA cycle}} \\
\midrule
CS & $\text{OAA} + {\color{gray}\text{AcCoA}} + {\color{gray}\text{H}_2\text{O}_\text{m}} \rightleftharpoons \text{CIT} + {\color{gray}\text{CoA}}$\\
ACO & $\text{CIT} \rightleftharpoons \text{ISOC}$\\
IDH & $\text{ISOC} + \text{NAD}\rightleftharpoons \alpha\text{KG} + \text{NADH} + {\color{gray} \text{CO}_2}$\\
KGDH & $\alpha\text{KG} + \text{NAD} + {\color{gray}\text{CoA}} \rightleftharpoons \text{SCoA} + \text{NADH} + {\color{gray} \text{CO}_2}$ \\
SL & $\text{SCoA} + \text{ADP}_\text{m} + {\color{gray}\text{Pi}_\text{m}} \rightleftharpoons \text{SUC} + \text{ATP}_\text{m} + {\color{gray}\text{CoA}}$\\
SDH & $\text{SUC} + {\color{gray} \text{CoQ}} \rightleftharpoons \text{FUM} + {\color{gray}\text{CoQH}_2}$\\
FH & $\text{FUM} + {\color{gray}\text{H}_2\text{O}_\text{m}} \rightleftharpoons \text{MAL}$\\
MDH & $\text{MAL} + \text{NAD} \rightleftharpoons \text{OAA} + \text{NADH}$\\
\midrule
\multicolumn{2}{l}{\textbf{Other exchange processes}} \\
\midrule
ANT & $\text{ATP}_\text{m} + \text{ADP}_\text{c} \rightleftharpoons \text{ATP}_\text{c} + \text{ADP}_\text{m}$\\
Hl & ${\color{gray}\text{H}^{+}_\text{c}} \rightleftharpoons {\color{gray}\text{H}^{+}_\text{m}}$\\
\midrule
\end{tabular}
\end{table}

\begin{table}
\renewcommand{\arraystretch}{2.}
\begin{fullwidth}
\caption{Fluxes of the system. $V_\mathrm{ref}$ is the volume of reference with respect to which each reaction rate, $J_\kappa$, and the corresponding entropy production rate, $\sigma_\kappa$, are normalized. Starting from the corresponding pseudoisomer concentrations, Magnus and Keizer estimate that $\left[\mathrm{ATP}^{4-}\right]_\mathrm{c}=0.05\, \left[\mathrm{ATP}\right]_\mathrm{c}$, $\left[\mathrm{ATP}^{4-}\right]_\mathrm{m}=0.05\, \left[\mathrm{ATP}\right]_\mathrm{m}$, $\left[\mathrm{ADP}^{3-}\right]_\mathrm{c}=0.45\,\left[\mathrm{ADP}\right]_\mathrm{c}$ and $\left[\mathrm{ADP}^{3-}\right]_\mathrm{m}=0.36\,\left[\mathrm{ADP}\right]_\mathrm{m}$.}
\label{tab:fluxes}
\centering
\resizebox{0.9\linewidth}{!}{
\begin{tabular}{llll}
\midrule
\textbf{Process} & $V_\mathrm{ref}$ & $J_\kappa~(\text{mM\,s}^{-1})$ & \textbf{Ref.} \\
\midrule
ACO & $V_m$ & $J_\mathrm{ACO}=k^\mathrm{ACO}_\mathrm{f} \left(\left[\mathrm{CIT}\right]_\mathrm{m} -\frac{\left[\mathrm{ISOC}\right]_\mathrm{m}}{K_\mathrm{ACO}}\right)$ & \cite{cortassa_integrated_2003} \\ 
ANT & $V_m$ & $J_\mathrm{ANT}=V^\mathrm{ANT}_{max} \frac{1 - \frac{\left[\mathrm{ATP}^{4-}\right]_\mathrm{c} \left[\mathrm{ADP}^{3-}\right]_\mathrm{m} }{\left[\mathrm{ATP}^{4-}\right]_\mathrm{m} \left[\mathrm{ADP}^{3-}\right]_\mathrm{c}} \mathrm{e}^{\frac{-\Delta \Psi}{RT}}}{\left(1 + \frac{\left[\mathrm{ATP}^{4-}\right]_\mathrm{c}}{\left[\mathrm{ADP}^{3-}\right]_\mathrm{c}} \mathrm{e}^{\frac{-f F \Delta \Psi}{RT}}\right) \left(1 + \frac{\left[\mathrm{ADP}^{3-}\right]_\mathrm{m}}{ \left[\mathrm{ATP}^{4-}\right]_\mathrm{m}}\right)}$ & \cite{magnus_minimal_1997} \\ 
CS & $V_m$ & $J_\mathrm{CS}=\frac{V_{max}^\mathrm{CS}}{1 + \frac{K_\mathrm{M,AcCoA}}{\left[\mathrm{AcCoA}\right]_\mathrm{m}} + \frac{K_\mathrm{M,OAA}^\mathrm{CS}}{\left[\mathrm{OAA}\right]_\mathrm{m}} \left(1 + \frac{\left[\mathrm{AcCoA}\right]_\mathrm{m}}{K_\mathrm{i,AcCoA}} \right) + \frac{K_\mathrm{s,AcCoA} K_\mathrm{M,OAA}^\mathrm{CS}}{(\left[\mathrm{OAA}\right]_\mathrm{m} \left[\mathrm{AcCoA}\right]_\mathrm{m}}}$ & \cite{dudycha_detailed_2000} \\ 
ERout & $V_c$ & $J_\mathrm{ERout}=\left(V^\mathrm{IP_3R}_{max} \frac{\left[\mathrm{IP}_3\right]^2}{\left[\mathrm{IP}_3\right]^2 + K_\mathrm{a,IP_3}^2} \frac{\left[\mathrm{Ca}^{2+}\right]_\mathrm{c}^2}{\left[\mathrm{Ca}^{2+}\right]_\mathrm{c}^2 + K_\mathrm{a,Cac}^2} \frac{K_\mathrm{i,Ca}^4}{K_\mathrm{i,Ca}^4 + \left[\mathrm{Ca}^{2+}\right]_\mathrm{c}^4}  + V^\mathrm{LEAK} \right) \left(\left[\mathrm{Ca}^{2+}\right]_\mathrm{ER} - \left[\mathrm{Ca}^{2+}\right]_\mathrm{c}\right)$ & \cite{komin_multiscale_2015} \\  
F1 & $V_m$ & $J_\mathrm{F1}=- \rho_\mathrm{f1}\frac{\left[p_a 10^{3\Delta \mathrm{pH}}+p_{c1}\mathrm{e}^{\frac{3F\Delta\Psi_B}{RT}}\right]A_\mathrm{F1}-p_a\mathrm{e}^{\frac{3F\Delta\Psi}{RT}}+p_{c2}A_\mathrm{F1}\mathrm{e}^{\frac{3F\Delta\Psi}{RT}}}{\left[1+p_1 A_\mathrm{F1}\right]\mathrm{e}^{\frac{3F\Delta\Psi_B}{RT}}+\left[p_2+p_3 A_\mathrm{F1}\right]\mathrm{e}^{\frac{3F\Delta\Psi}{RT}}}$ & \cite{magnus_minimal_1997} \\  
& & with $A_\mathrm{F1} = K_\mathrm{F1}  \frac{\left[\mathrm{ATP}\right]_\mathrm{m}}{\left[\mathrm{ADP}\right]_\mathrm{m} \left[\mathrm{Pi}\right]_\mathrm{m}}$ & \\
FH & $V_m$ & $J_\mathrm{FH}=k^\mathrm{FH}_\mathrm{f} \left(\left[\mathrm{FUM}\right]_\mathrm{m} -\frac{\left[\mathrm{MAL}\right]_\mathrm{m}}{K_\mathrm{FH}}\right)$ & \cite{cortassa_integrated_2003} \\  
Hl & $V_m$ & $J_\mathrm{Hl}=g_\mathrm{H} \left(\Delta \Psi - 2.303 \frac{RT}{F} \Delta \mathrm{pH}\right)$ & \cite{magnus_minimal_1997} \\ 
Hyd & $V_c$ & $J_\mathrm{Hyd}=k_\mathrm{Hyd} \frac{\left[\mathrm{ATP}\right]_\mathrm{c}}{\left[\mathrm{ATP}\right]_\mathrm{c}+K_\mathrm{M,ATPc}}$ & \cite{wacquier_interplay_2016} \\ 
IDH & $V_m$ & $J_\mathrm{IDH}=\frac{V_{max}^\mathrm{IDH}}{1+\frac{\left[\mathrm{H}\right]_\mathrm{m}}{k_\mathrm{h,1}}+\frac{k_\mathrm{h,2}}{\left[\mathrm{H}\right]_\mathrm{m}} + \frac{\left(\frac{K_\mathrm{M, ISOC}}{\left[\mathrm{ISOC}\right]_\mathrm{m}}\right)^{n_i}}{\left(1+\frac{\left[\mathrm{ADP}\right]_\mathrm{m}}{K_\mathrm{a,ADP}}\right)\left(1+\frac{\left[\mathrm{Ca}^{2+}\right]_\mathrm{m}}{K_\mathrm{a,Cam}}\right)}+\frac{K_\mathrm{M,NAD}^\mathrm{IDH}}{\left[\mathrm{NAD}\right]_\mathrm{m}}\left(1 + \frac{\left[\mathrm{NADH}\right]_\mathrm{m}}{K_\mathrm{i,NADH}}\right) + \frac{\left(\frac{K_\mathrm{M, ISOC}}{\left[\mathrm{ISOC}\right]_\mathrm{m}}\right)^{n_i} \frac{K_\mathrm{M,NAD}^\mathrm{IDH}}{\left[\mathrm{NAD}\right]_\mathrm{m}}\left(1 + \frac{\left[\mathrm{NADH}\right]_\mathrm{m}}{K_\mathrm{i,NADH}}\right)}{\left(1+\frac{\left[\mathrm{ADP}\right]_\mathrm{m}}{K_\mathrm{a,ADP}}\right)\left(1+\frac{\left[\mathrm{Ca}^{2+}\right]_\mathrm{m}}{K_\mathrm{a,Cam}}\right)}}$ & \cite{cortassa_integrated_2003} \\  
KGDH & $V_m$ & $J_\mathrm{KGDH}=\frac{V_{max}^\mathrm{KGDH}}{1+\frac{\frac{K_\mathrm{M,\alpha KG}}{\left[\mathrm{\alpha KG}\right]_\mathrm{m}} \left(\frac{K_\mathrm{M,NAD}^\mathrm{KGDH}}{\left[\mathrm{NAD}\right]_\mathrm{m}}\right)^{n_{\mathrm{\alpha KG}}}}{\left(1+\frac{\left[\mathrm{Mg}^{2+}\right]_\mathrm{m}}{K_\mathrm{D,Mg}}\right)\left(1+\frac{\left[\mathrm{Ca}^{2+}\right]_\mathrm{m}}{K_\mathrm{D,Ca}}\right)}}$
& \cite{dudycha_detailed_2000} \\  
MDH & $V_m$ & $J_\mathrm{MDH}=V^\mathrm{MDH}_{max} \frac{\left[\mathrm{MAL}\right]_\mathrm{m} \left[\mathrm{NAD}\right]_\mathrm{m} - \frac{\left[\mathrm{OAA}\right]_\mathrm{m} \left[\mathrm{NADH}\right]_\mathrm{m}}{K_\mathrm{MDH}}}{\left(1+\frac{\left[\mathrm{MAL}\right]_\mathrm{m}}{K_\mathrm{M,MAL}}\right)\left(1+\frac{\left[\mathrm{NAD}\right]_\mathrm{m}}{K_\mathrm{M,NAD}^\mathrm{MDH}}\right)+\left(1+\frac{\left[\mathrm{OAA}\right]_\mathrm{m}}{K_\mathrm{M,OAA}^\mathrm{MDH}}\right)\left(1+\frac{\left[\mathrm{NADH}\right]_\mathrm{m}}{K_\mathrm{M,NADH}}\right)-1}$ & \cite{berndt_physiology-based_2015} \\  
NCX & $V_m$ & $J_\mathrm{NCX}=V^\mathrm{NCX}_{max}\frac{\mathrm{e}^{\frac{bF\left(\Delta\Psi-\Delta\Psi^*\right)}{RT}}}{\left(1+\frac{K_\mathrm{M,Na}}{\left[\mathrm{Na}^{+}\right]_\mathrm{c}}\right)^n\left(1+\frac{K_\mathrm{M,Ca}}{\left[\mathrm{Ca}^{2+}\right]_\mathrm{m}}\right)}$ & \cite{magnus_minimal_1997}  \\  
Ox & $V_m$ & $J_\mathrm{Ox}=\frac{1}{2} \rho_\mathrm{res}\frac{\left[r_a 10^{6\Delta \mathrm{pH}}+r_{c1}\mathrm{e}^{\frac{6F\Delta\Psi_B}{RT}}\right]A_\mathrm{res}-r_a\mathrm{e}^{\frac{g6F\Delta\Psi}{RT}}+r_{c2}A_\mathrm{res}\mathrm{e}^{\frac{g6F\Delta\Psi}{RT}}}{\left[1+r_1 A_\mathrm{res}\right]\mathrm{e}^{\frac{6F\Delta\Psi_B}{RT}}+\left[r_2+r_3 A_\mathrm{res}\right]\mathrm{e}^{\frac{g6F\Delta\Psi}{RT}}}$ & \cite{magnus_minimal_1997} \\  
& & with $A_\mathrm{res} = K_\mathrm{res}  \sqrt{\frac{\left[\mathrm{NADH}\right]_\mathrm{m}}{\left[\mathrm{NAD}\right]_\mathrm{m}}}$ & \\
SDH & $V_m$ & $J_\mathrm{SDH}=\frac{V_{max}^\mathrm{SDH}}{1+\frac{K_\mathrm{M,SUC}}{\left[\mathrm{SUC}\right]_\mathrm{m}}\left(1+\frac{\left[\mathrm{OAA}\right]_\mathrm{m}}{K_\mathrm{i,OAA}}\right)\left(1+\frac{\left[\mathrm{FUM}\right]_\mathrm{m}}{K_\mathrm{i,FUM}}\right)}$ & \cite{cortassa_integrated_2003} \\  
SERCA & $V_c$ & $J_\mathrm{SERCA}=V_{max}^\mathrm{SERCA} \frac{\left[\mathrm{Ca}^{2+}\right]_\mathrm{c}^{2}}{\left[\mathrm{Ca}^{2+}\right]_\mathrm{c}^{2} + K_\mathrm{Ca}^2} \frac{\left[\mathrm{ATP}\right]_\mathrm{c}}{\left[\mathrm{ATP}\right]_\mathrm{c}+K_\mathrm{ATPc}}$ & \cite{wacquier_interplay_2016} \\  
SL & $V_m$ & $J_\mathrm{SL}=k^\mathrm{SL}_\mathrm{f} \left(\left[\mathrm{SCoA}\right]_\mathrm{m} \left[\mathrm{ADP}\right]_\mathrm{m} \left[\mathrm{Pi}\right]_\mathrm{m} -\frac{\left[\mathrm{SUC}\right]_\mathrm{m} \left[\mathrm{ATP}\right]_\mathrm{m} \left[\mathrm{CoA}\right]_\mathrm{m}}{K_\mathrm{SL}}\right)$ & \cite{wei_mitochondrial_2011} \\  
UNI & $V_m$ & $J_\mathrm{UNI}=V_{max}^\mathrm{UNI} \frac{2F \left(\Delta \Psi-\Delta \Psi^*\right)}{RT\left(1-\mathrm{e}^{-\frac{2F\left(\Delta \Psi-\Delta \Psi^*\right)}{RT}}\right)} \frac{\frac{\left[\mathrm{Ca}^{2+}\right]_\mathrm{c}}{K_\mathrm{trans}}\left(1+\frac{\left[\mathrm{Ca}^{2+}\right]_\mathrm{c}}{K_\mathrm{trans}}\right)^3}{\left(1+\frac{\left[\mathrm{Ca}^{2+}\right]_\mathrm{c}}{K_\mathrm{trans}}\right)^4+\frac{L}{\left(1+\frac{\left[\mathrm{Ca}^{2+}\right]_\mathrm{c}}{K_\mathrm{act}}\right)^{n_a}}}$ & \cite{magnus_minimal_1997} \\
\midrule
\end{tabular}}
\end{fullwidth}
\end{table}

\begin{table}
\renewcommand{\arraystretch}{2.}
\caption{Forces of the system. Transformed Gibbs free energies of reaction ($\Delta_r G_\kappa$) associated to each process of the system. The indices $c$ and $m$ associated to $\Delta_r G'^\circ_\kappa$ indicate that this thermodynamic quantity is evaluated at cytosolic and mitochondrial pH, that is, $\mathrm{pH}=7.2$ and $\mathrm{pH}=8.0$, respectively. The value of $\Delta_r G'^\circ_\kappa$, which also accounts for physiological ionic strength ($I=0.12$ M \cite{robinson_em_2006}) and $\left[\mathrm{Mg^{2+}}\right]_\mathrm{m}$ (pMg=3.4), was retrieved for each relevant process \textit{via} Equilibrator \cite{flamholz_equilibratorbiochemical_2012}. Starting from the corresponding pseudoisomer concentrations, Magnus and Keizer estimate that $\left[\mathrm{ATP}^{4-}\right]_\mathrm{c}=0.05\, \left[\mathrm{ATP}\right]_\mathrm{c}$, $\left[\mathrm{ATP}^{4-}\right]_\mathrm{m}=0.05\, \left[\mathrm{ATP}\right]_\mathrm{m}$, $\left[\mathrm{ADP}^{3-}\right]_\mathrm{c}=0.45\,\left[\mathrm{ADP}\right]_\mathrm{c}$ and $\left[\mathrm{ADP}^{3-}\right]_\mathrm{m}=0.36\,\left[\mathrm{ADP}\right]_\mathrm{m}$.}
\label{tab:forces}
\centering
\resizebox{\linewidth}{!}{
\begin{tabular}{lll}
\midrule
\textbf{Process} & $\Delta_r G'_\kappa~(\mathrm{J\,mol}^{-1})$ &  \\ \midrule
ACO & $\Delta_r G'_\mathrm{ACO,m} = \Delta_r G'^\circ_\mathrm{ACO,m} + RT \ln \frac{\left[\mathrm{ISOC}\right]_\mathrm{m}}{\left[\mathrm{CIT}\right]_\mathrm{m}}$ & $\Delta_r G'^\circ_\mathrm{ACO,m} = 6700\, \mathrm{J\, mol^{-1}}$ \\ 
ANT & $\Delta_r G'_\mathrm{ANT,m} =RT \ln \frac{\left[\mathrm{ATP}^{4-}\right]_\mathrm{c} \left[\mathrm{ADP}^{3-}\right]_\mathrm{m}}{\left[\mathrm{ATP}^{4-}\right]_\mathrm{m} \left[\mathrm{ADP}^{3-}\right]_\mathrm{c}} - F \Delta \Psi$ & \\ 
CS & $\Delta_r G'_\mathrm{CS,m} =\Delta_r G'^\circ_\mathrm{CS,m} + RT \ln \frac{\left[\mathrm{CIT}\right]_\mathrm{m} \left[\mathrm{CoA}\right]_\mathrm{m}}{\left[\mathrm{OAA}\right]_\mathrm{m} \left[\mathrm{AcCoA}\right]_\mathrm{m}}$ & $\Delta_r G'^\circ_\mathrm{CS,m} = -41200\, \mathrm{J\, mol^{-1}}$ \\ 
ERout & $\Delta_r G'_\mathrm{ERout,c} =RT \ln \frac{\left[\mathrm{Ca}^{2+}\right]_\mathrm{c}}{\left[\mathrm{Ca}^{2+}\right]_\mathrm{ER}}$ & \\  
F1 & $\Delta_r G'_\mathrm{F1,m} =- \Delta_r G'^\circ_\mathrm{Hyd,m} + RT \ln \frac{\left[\mathrm{H}\right]_\mathrm{m}^3 \left[\mathrm{ATP}\right]_\mathrm{m}}{\left[\mathrm{H}\right]_\mathrm{c}^3 \left[\mathrm{ADP}\right]_\mathrm{m} \left[\mathrm{Pi}\right]_\mathrm{m}} - 3 F \Delta \Psi $ & $\Delta_r G'^\circ_\mathrm{Hyd,m} = -32200\, \mathrm{J\, mol^{-1}}$ \\  
FH & $\Delta_r G'_\mathrm{FH,m} =\Delta_r G'^\circ_\mathrm{FH,m} + RT \ln \frac{\left[\mathrm{MAL}\right]_\mathrm{m}}{\left[\mathrm{FUM}\right]_\mathrm{m}}$ & $\Delta_r G'^\circ_\mathrm{FH,m} = -3400 \, \mathrm{J\, mol^{-1}}$  \\  
Hl & $\Delta_r G'_\mathrm{Hl,m} =RT \ln \frac{\left[\mathrm{H}\right]_\mathrm{m}}{\left[\mathrm{H}\right]_\mathrm{c}} - F \Delta \Psi$ & \\  
Hyd & $\Delta_r G'_\mathrm{Hyd,c} =\Delta_r G'^\circ_\mathrm{Hyd,c} + RT \ln \frac{\left[\mathrm{ADP}\right]_\mathrm{c} \left[\mathrm{Pi}\right]_\mathrm{c}}{\left[\mathrm{ATP}\right]_\mathrm{c}}$ & $\Delta_r G'^\circ_\mathrm{Hyd, c} = -28300 \, \mathrm{J\, mol^{-1}}$ \\  
IDH & $\Delta_r G'_\mathrm{IDH,m} =\Delta_r G'^\circ_\mathrm{IDH,m} + RT \ln \frac{\left[\mathrm{\alpha KG}\right]_\mathrm{m} \left[\mathrm{CO_2}\right]_\mathrm{m} \left[\mathrm{NADH}\right]_\mathrm{m}}{\left[\mathrm{ISOC}\right]_\mathrm{m} \left[\mathrm{NAD}\right]_\mathrm{m}}$ & $\Delta_r G'^\circ_\mathrm{IDH,m} = 5100 \, \mathrm{J\, mol^{-1}}$ \\  
KGDH & $\Delta_r G'_\mathrm{KGDH,m} =\Delta_r G'^\circ_\mathrm{KGDH,m} + RT \ln \frac{\left[\mathrm{SCoA}\right]_\mathrm{m} \left[\mathrm{NADH}\right]_\mathrm{m} \left[\mathrm{CO_2}\right]_\mathrm{m}}{\left[\mathrm{\alpha KG}\right]_\mathrm{m} \left[\mathrm{NAD}\right]_\mathrm{m} \left[\mathrm{CoA}\right]_\mathrm{m}}$ & $\Delta_r G'^\circ_\mathrm{KGDH,m} = -27600 \, \mathrm{J\, mol^{-1}}$ \\  
MDH & $\Delta_r G'_\mathrm{MDH,m} =\Delta_r G'^\circ_\mathrm{MDH,m} + RT \ln \frac{\left[\mathrm{OAA}\right]_\mathrm{m} \left[\mathrm{NADH}\right]_\mathrm{m}}{\left[\mathrm{NAD}\right]_\mathrm{m} \left[\mathrm{MAL}\right]_\mathrm{m}}$ & $\Delta_r G'^\circ_\mathrm{MDH,m} = 24200 \, \mathrm{J\, mol^{-1}}$ \\  
NCX & $\Delta_r G'_\mathrm{NCX,m} =RT \ln \frac{\left[\mathrm{Ca}^{2+}\right]_\mathrm{c} \left[\mathrm{Na}\right]_\mathrm{m}^3}{\left[\mathrm{Ca}^{2+}\right]_\mathrm{m} \left[\mathrm{Na}\right]_\mathrm{c}^3} - F \Delta \Psi$ & \\  
Ox & $\Delta_r G'_\mathrm{Ox,m} =\Delta_r G'^\circ_\mathrm{Ox,m} + RT \ln \frac{\left[\mathrm{H}\right]_\mathrm{c}^{10} \left[\mathrm{NAD}\right]_\mathrm{m}}{\left[\mathrm{H}\right]_\mathrm{m}^{10} \left[\mathrm{NADH}\right]_\mathrm{m} \left[\mathrm{O_2}\right]_\mathrm{m}^{0.5}} + 10 F \Delta \Psi$ & $\Delta_r G'^\circ_\mathrm{Ox,m} = -225300 \, \mathrm{J\, mol^{-1}}$ \\  
SDH & $\Delta_r G'_\mathrm{SDH,m} =\Delta_r G'^\circ_\mathrm{SDH,m} + RT \ln \frac{\left[\mathrm{FUM}\right]_\mathrm{m} \left[\mathrm{CoQH_2}\right]_\mathrm{m}}{\left[\mathrm{SUC}\right]_\mathrm{m} \left[\mathrm{CoQ}\right]_\mathrm{m}}$ & $\Delta_r G'^\circ_\mathrm{SDH,m} = -24200 \, \mathrm{J\, mol^{-1}}$ \\  
SERCA & $\Delta_r G'_\mathrm{SERCA,c} =\Delta_r G'^\circ_\mathrm{Hyd,c} + RT \ln \frac{\left[\mathrm{ADP}\right]_\mathrm{c} \left[\mathrm{Pi}\right]_\mathrm{c} \left[\mathrm{Ca}^{2+}\right]_\mathrm{ER}^2}{\left[\mathrm{ATP}\right]_\mathrm{c} \left[\mathrm{Ca}^{2+}\right]_\mathrm{c}^2}$ & $\Delta_r G'^\circ_\mathrm{Hyd, c} = -28300 \, \mathrm{J\, mol^{-1}}$ \\  
SL & $\Delta_r G'_\mathrm{SL,m} =\Delta_r G'^\circ_\mathrm{SL,m} + RT \ln \frac{\left[\mathrm{SUC}\right]_\mathrm{m} \left[\mathrm{CoA}\right]_\mathrm{m} \left[\mathrm{ATP}\right]_\mathrm{m}}{\left[\mathrm{SCoA}\right]_\mathrm{m} \left[\mathrm{ADP}\right]_\mathrm{m} \left[\mathrm{Pi}\right]_\mathrm{m}}$ & $\Delta_r G'^\circ_\mathrm{SL,m} = 800 \, \mathrm{J\, mol^{-1}}$ \\  
UNI & $\Delta_r G'_\mathrm{UNI,m} =RT \ln \frac{\left[\mathrm{Ca}^{2+}\right]_\mathrm{m}}{\left[\mathrm{Ca}^{2+}\right]_\mathrm{c}} - 2 F \Delta \Psi$ & \\
\midrule
\end{tabular}}
\end{table}

\renewcommand{\arraystretch}{1.}
\begin{table}
    \begin{fullwidth}
    \caption{Reference parameter values.}
    \label{tab:param}
    \centering
    \resizebox{\linewidth}{!}{
    \begin{tabular}{llll}
        \hline
        \textbf{Parameter} & \textbf{Definition} & \textbf{Value (units)} & \textbf{Ref.} \\ \hline
        $\alpha$ & Ratio between ER and cytosol volumes & 0.10 & \cite{wacquier_interplay_2016} \\
        $A_\mathrm{tot}$ & Total concentration of cytosolic adenine nucleotides & 3 mM & \cite{moein_dissecting_2017} \\
        $A_\mathrm{m,tot}$ & Total concentration of mitochondrial adenine nucleotides & 15 mM & \cite{magnus_minimal_1997}\\
        $b$ & Dependence of electrogenic Na\textsuperscript{+}/Ca\textsuperscript{2+} exchanger on $\Delta \Psi$ & 0.5 & \cite{magnus_minimal_1997} \\
        $C_m$ & Mitochondrial membrane capacitance & $1.812 \times 10^{-3}$ mM mV\textsuperscript{-1} & \cite{cortassa_integrated_2003} \\
        $\left[\mathrm{CO_2}\right]$ & Total CO\textsubscript{2} concentration in mitochondrial matrix & 21.4 mM &  \cite{wu_computer_2007} \\
        $\left[\mathrm{CoA}\right]$ & CoA concentration in mitochondrial matrix & 0.02 mM & \cite{cortassa_integrated_2003} \\
        $\left[\mathrm{CoQ}\right]$ & CoQ concentration in mitochondrial matrix & 0.97 mM & \cite{wu_computer_2007} \\
        $\left[\mathrm{CoQH_2}\right]$ & CoQ\textsubscript{2} concentration in mitochondrial matrix & 0.38 mM & \cite{wu_computer_2007} \\
        $c_\mathrm{tot}$ & Total free Ca\textsuperscript{2+} concentration of the cell normalized by $V_c$ & 1500 $\mu$M & This work \\
        $c_\mathrm{Ktot}$ & Total concentration of TCA cycle intermediates & 1 mM & \cite{cortassa_integrated_2003} \\
        $\delta$ & Ratio between mitochondrial matrix and cytosol volumes & 0.15 & \cite{siess_subcellular_1976, lund_matrix_1987}\\
        $\Delta \mathrm{pH}$ & pH difference between cytosol and mitochondrial matrix ($\mathrm{pH_c} - \mathrm{pH_m}$) & -0.80 & \cite{buckler_application_1990, casey_sensors_2010}\\
        $\Delta \Psi^*$ & Membrane potential offset for Ca\textsuperscript{2+} transport & 91 mV & \cite{magnus_minimal_1997} \\
        $\Delta\Psi_B$ & Total phase boundary potential & 50 mV & \cite{magnus_minimal_1997} \\
        $F$ & Faraday constant & 96.485 kC mol\textsuperscript{-1} & \\
        $f$ & Fraction of $\Delta \Psi$ responsible for the behavior of ANT in energized mitochondria & 0.5 & \cite{magnus_minimal_1997} \\
        $f_c$ & Fraction of free cytosolic Ca\textsuperscript{2+} & 0.01 & \cite{wacquier_interplay_2016} \\
        $f_e$ & Fraction of free Ca\textsuperscript{2+} in the ER & 0.01 & \cite{wacquier_interplay_2016} \\
        $f_m$ & Fraction of free mitochondrial Ca\textsuperscript{2+} & 0.0003 & \cite{magnus_minimal_1997} \\
        $\gamma$ & Conversion factor between mM and $\mu$M & 1000 $\mu$M mM\textsuperscript{-1} & \\
        $g$ & Fitting factor for voltage in respiration rate & 0.85 & \cite{magnus_minimal_1997} \\
        $g_\mathrm{H}$ & Ionic conductance of the mitochondrial inner membrane & $10^{-5}$ mM mV\textsuperscript{-1} s\textsuperscript{-1} & \cite{cortassa_integrated_2003} \\
        $\left[\mathrm{H^+}\right]_\mathrm{c}$ & Cytosolic proton concentration & $6.31 \times 10^{-5}$ mM & \cite{buckler_application_1990, casey_sensors_2010}\\
        $\left[\mathrm{H^+}\right]_\mathrm{m}$ & Concentration of proton in the mitochondrial matrix & $10^{-5}$ mM & \cite{buckler_application_1990, casey_sensors_2010}\\
        $K_\mathrm{a,Cac}$ & Activation constant of IP\textsubscript{3}Rs for cytosolic Ca\textsuperscript{2+} & 0.60 $\mu$M & This work \\
        $K_\mathrm{ACO}$ & Equilibrium constant of ACO & 0.067 & \cite{flamholz_equilibratorbiochemical_2012, berndt_physiology-based_2015}\\
        $K_\mathrm{act}$ & Dissociation constant of mitochondrial uniporter for activating Ca\textsuperscript{2+} & 0.38 & \cite{magnus_minimal_1997}\\
        $K_\mathrm{ATPc}$ & Dissociation constant of SERCA for cytosolic ATP & 0.05 mM & \cite{scofano_substrate_1979, moein_dissecting_2017}\\
        $K_\mathrm{a,ADP}$ & Activation constant of IDH for ADP\textsubscript{m} & 0.062 mM & \cite{dudycha_detailed_2000, cortassa_integrated_2003}\\
        $K_\mathrm{a,Cam}$ & Activation constant of IDH for mitochondrial Ca\textsuperscript{2+} & 1.41 $\mu$M & \cite{cortassa_integrated_2003}\\
        $K_\mathrm{a,IP_3}$ & Activation constant of IP\textsubscript{3}Rs for IP\textsubscript{3} & 1.00 $\mu$M & \cite{dupont_simulations_1997, wacquier_interplay_2016} \\
        $K_\mathrm{Ca}$ & Dissociation constant of SERCA for Ca\textsuperscript{2+} & 0.35 $\mu$M & \cite{dupont_simulations_1997, wacquier_interplay_2016} \\
        $K_\mathrm{D,Ca}$ & Dissociation constant of KGDH for mitochondrial Ca\textsuperscript{2+} & 1.27 $\mu$M & \cite{dudycha_detailed_2000, cortassa_integrated_2003}\\
        $K_\mathrm{D,Mg}$ & Dissociation constant of KGDH for mitochondrial Mg\textsuperscript{2+} & 0.0308 mM & \cite{cortassa_integrated_2003} \\
        $K_\mathrm{F1}$ & Equilibrium constant for ATP hydrolysis in mitochondrial matrix & $1.71 \times 10^{6}$ & \cite{pietrobon_flow-force_1985, cortassa_integrated_2003} \\
        $K_\mathrm{FH}$ & Equilibrium constant for FH & 3.942 & \cite{flamholz_equilibratorbiochemical_2012} \\
        $k^\mathrm{ACO}_\mathrm{f}$ & Forward rate constant of ACO & 12.5 s\textsuperscript{-1} & \cite{cortassa_integrated_2003} \\
        $k^\mathrm{FH}_\mathrm{f}$ & Forward rate constant of FH & 8.3 s\textsuperscript{-1} & This work \\
        $k^\mathrm{SL}_\mathrm{f}$ & Forward rate constant of SL & 0.127 mM\textsuperscript{-2} s\textsuperscript{-1} & \cite{cortassa_integrated_2003} \\
        $k_\mathrm{h,1}$ & First ionization constant of IDH & $8.1 \times 10^{-5}$ mM & \cite{dudycha_detailed_2000, cortassa_integrated_2003}\\
        $k_\mathrm{h,2}$ & Second ionization constant of IDH & $5.98 \times 10^{-5}$ mM & \cite{dudycha_detailed_2000, cortassa_integrated_2003} \\
        $k_\mathrm{Hyd}$ & Hydrolysis rate of ATP\textsubscript{c} due to cellular activity & $2 \times 10^{-2}$ mM s\textsuperscript{-1} & This work \\
        $K_\mathrm{i,AcCoA}$ & Inhibition constant of CS for AcCoA & $3.7068 \times 10^{-2}$ mM & \cite{dudycha_detailed_2000}\\
        $K_\mathrm{i,Ca}$ & Inhibition constant of IP\textsubscript{3}Rs for cytosolic Ca\textsuperscript{2+} & 1.00 $\mu$M & This work \\
        $K_\mathrm{i,FUM}$ & Inhibition constant of SDH for fumarate & 1.3 mM & \cite{cortassa_integrated_2003} \\
        $K_\mathrm{i,OAA}$ & Inhibition constant of SDH for oxaloacetate & 0.15 mM & \cite{cortassa_integrated_2003} \\
        $K_\mathrm{i,NADH}$ & Inhibition constant of IDH for NADH & 0.19 mM & \cite{cortassa_integrated_2003} \\
        $K_\mathrm{M,AcCoA}$ & Michaelis constant of CS for acetyl-CoA & $1.2614 \times 10^{-2}$ mM & \cite{dudycha_detailed_2000, cortassa_integrated_2003}\\
        $K_\mathrm{M,\alpha KG}$ & Michaelis constant of KGDH for $\alpha$-ketoglutarate & 1.94 mM & \cite{cortassa_integrated_2003} \\
        $K_\mathrm{M,ATPc}$ & Michaelis constant for ATP\textsubscript{c} hydrolysis due to cellular activity & 1 mM & \cite{wacquier_interplay_2016} \\
        $K_\mathrm{M,Ca}$ & Michaelis constant of Na\textsuperscript{+}/Ca\textsuperscript{2+} exchanger for Ca\textsuperscript{2+} & 0.375 $\mu$M & \cite{cortassa_integrated_2003} \\
        $K_\mathrm{M, ISOC}$ & Michaelis constant of IDH for isocitrate & $1.52$ mM & \cite{dudycha_detailed_2000, cortassa_integrated_2003}\\
        $K_\mathrm{M,MAL}$ & Michaelis constant of MDH for malate & 0.145 mM & \cite{berndt_physiology-based_2015}\\
        $K_\mathrm{M,Na}$ & Michaelis constant of Na\textsuperscript{+}/Ca\textsuperscript{2+} exchanger for Na\textsuperscript{+} & 9.4 mM & \cite{magnus_minimal_1997} \\
        $K_\mathrm{M,NAD}^\mathrm{IDH}$ & Michaelis constant of IDH for NAD & 0.923 mM & \cite{dudycha_detailed_2000, cortassa_integrated_2003} \\
        $K_\mathrm{M,NAD}^\mathrm{KGDH}$ & Michaelis constant of KGDH for NAD & $3.87 \times 10^{-2}$ mM & This work \\
        $K_\mathrm{M,NAD}^\mathrm{MDH}$ & Michaelis constant of MDH for NAD & 0.06 mM & \cite{berndt_physiology-based_2015}\\
        $K_\mathrm{M,NADH}$ & Michaelis constant of MDH for NADH & 0.044 mM & \cite{cortassa_integrated_2003}\\
         & & & \\
        \hline
    \end{tabular}
    }
  \end{fullwidth}
\end{table}

\setcounter{table}{3}

\renewcommand{\arraystretch}{1.}
\begin{table}
    \begin{fullwidth}
    \caption{Reference parameter values (continued).}
    \centering
    \resizebox{\linewidth}{!}{
    \begin{tabular}{llll}
        \hline
        \textbf{Parameter} & \textbf{Definition} & \textbf{Value (units)} & \textbf{Ref.} \\ \hline
        $K_\mathrm{M,OAA}^\mathrm{CS}$ & Michaelis constant of CS for oxaloacetate & $5 \times 10^{-3}$ mM & \cite{matsuoka_kinetic_1973, kurz_catalytic_1995, berndt_physiology-based_2015} \\
        $K_\mathrm{M,OAA}^\mathrm{MDH}$ & Michaelis constant of MDH for oxaloacetate & 0.017 mM & \cite{berndt_physiology-based_2015}\\
        $K_\mathrm{M,SUC}$ & Michaelis constant of SDH for succinate & $3 \times 10^{-2}$ mM & \cite{cortassa_integrated_2003} \\
        $K_\mathrm{MDH}$ & Equilibrium constant of MDH & $2.756 \times 10^{-5}$ & \cite{flamholz_equilibratorbiochemical_2012}\\
        $K_\mathrm{res}$ & Equilibium constant of O\textsubscript{2} reduction by NADH in mitochondrial matrix & $1.35 \times 10^{18}$ & \cite{magnus_minimal_1997} \\
        $K_\mathrm{s,AcCoA}$ & Other binding constant of citrate synthase for AcCoA & $8.0749 \times 10^{-2}$ mM & \cite{dudycha_detailed_2000}\\
        $K_\mathrm{SL}$ & Equilibrium constant for SL & 0.724 & \cite{flamholz_equilibratorbiochemical_2012} \\
        $K_\mathrm{trans}$ & Dissociation constant of mitochondrial uniporter for translocated Ca\textsuperscript{2+} & 19 $\mu$M & \cite{magnus_model_1998-c} \\
        $L$ & Equilibrium constant for mitochondrial uniporter conformations & 110 & \cite{magnus_model_1998-c} \\
        $\left[\mathrm{Mg^{2+}}\right]_\mathrm{m}$ & Mg concentration in the mitochondrial matric & 0.4 mM & \cite{cortassa_integrated_2003} \\
        $n$ & Number of Na\textsuperscript{+} binding to electrogenic Na\textsuperscript{+}/Ca\textsuperscript{2+} exchanger & 3 & \cite{magnus_minimal_1997} \\
        $n_a$ & Mitochondrial uniporter activation cooperativity & 2.8 & \cite{magnus_minimal_1997} \\
        $\left[\mathrm{Na^+}\right]_\mathrm{c}$ & Cytosolic Na\textsuperscript{+} concentration & 10 mM & \cite{cortassa_integrated_2003}\\
        $\left[\mathrm{Na^+}\right]_\mathrm{m}$ & Mitochondrial Na\textsuperscript{+} concentration & 5 mM & \cite{donoso_fluorescence_1992}\\
        $n_{\mathrm{\alpha KG}}$ & Hill coefficient of KGDH for $\alpha$KG  & 1.2 & \cite{cortassa_integrated_2003}\\
        $n_i$ & Hill coefficient of IDH for isocitrate & 2 & \cite{wei_mitochondrial_2011} \\
        $N_\mathrm{tot}$ & Total concentration of mitochondrial pyridine nucleotides & 0.8 mM & This work \\
        $\left[\mathrm{O_2}\right]$ & O\textsubscript{2} concentration in mitochondrial matrix & $2.6 \times 10^{-5}$ M & \cite{beard_biophysical_2005}\\
        $p_1$ & Combination of elementary kinetic constants for the 6-state ATPase model & $1.346 \times 10^{-8}$ & \cite{magnus_minimal_1997}\\
        $p_2$ & Combination of elementary kinetic constants for the 6-state ATPase model & $7.739 \times 10^{-7}$ & \cite{magnus_minimal_1997}\\
        $p_3$ & Combination of elementary kinetic constants for the 6-state ATPase model & $6.65 \times 10^{-15}$ & \cite{magnus_minimal_1997}\\
        $p_a$ & Combination of elementary kinetic constants for the 6-state ATPase model & $1.656 \times 10^{-5}$ s\textsuperscript{-1} & \cite{magnus_minimal_1997}\\
        $p_{c1}$ & Combination of elementary kinetic constants for the 6-state ATPase model & $9.651 \times 10^{-14}$ s\textsuperscript{-1} & \cite{magnus_minimal_1997}\\
        $p_{c2}$ & Combination of elementary kinetic constants for the 6-state ATPase model & $4.845 \times 10^{-19}$ s\textsuperscript{-1} & \cite{magnus_minimal_1997}\\
        $\left[\mathrm{P_i}\right]_\mathrm{c}$ & Inorganic phosphate concentration in cytosol & 1 mM & \cite{bevington_study_1986}\\
        $\left[\mathrm{P_i}\right]_\mathrm{m}$ & Inorganic phosphate concentration in mitochondrial matrix & 20 mM & \cite{magnus_minimal_1997} \\
        $R$ & Gas constant & 8.314 J mol\textsuperscript{-1} K\textsuperscript{-1} & \\
        $\rho_\mathrm{f1}$ & Density of ATPase pumps & 1.5 & This work\\
        $\rho_\mathrm{res}$ & Density of H\textsuperscript{+} pumps in mitochondrial membrane & 1.00 & This work \\
        $r_1$ & Combination of elementary kinetic constants for the 6-state respiration model & $2.077 \times 10^{-18}$ & \cite{magnus_minimal_1997} \\
        $r_2$ & Combination of elementary kinetic constants for the 6-state respiration model & $1.728 \times 10^{-9}$ & \cite{magnus_minimal_1997} \\
        $r_3$ & Combination of elementary kinetic constants for the 6-state respiration model & $1.059\times 10^{-26}$ & \cite{magnus_minimal_1997} \\
        $r_a$ & Combination of elementary kinetic constants for the 6-state respiration model & $6.394\times 10^{-10}$ s\textsuperscript{-1} & \cite{magnus_minimal_1997} \\
        $r_{c1}$ & Combination of elementary kinetic constants for the 6-state respiration model & $2.656\times 10^{-19}$ s\textsuperscript{-1} & \cite{magnus_minimal_1997} \\
        $r_{c2}$ & Combination of elementary kinetic constants for the 6-state respiration model & $8.632\times 10^{-27}$ s\textsuperscript{-1} & \cite{magnus_minimal_1997} \\
        $T$ & Temperature & 310 K & \cite{cortassa_integrated_2003}\\
         $V^\mathrm{ANT}_{max}$ & Limiting rate of adenine nucleotide translocator (ANT) & 15 mM s\textsuperscript{-1} & \cite{cortassa_integrated_2003} \\
        $V_{max}^\mathrm{CS}$ & Limiting rate of CS & 52 mM s\textsuperscript{-1} & This work\\
        $V_{max}^\mathrm{IDH}$ & Limiting rate of IDH & 0.15 mM s\textsuperscript{-1} & This work\\
        $V^\mathrm{IP_3R}_{max}$ & Limiting release rate of Ca\textsuperscript{2+} through IP\textsubscript{3}Rs & 15 s\textsuperscript{-1} & This work \\
        $V_{max}^\mathrm{KGDH}$ & Limiting rate of KGDH & 5 mM s\textsuperscript{-1} & This work \\
        $V^\mathrm{LEAK}$ & Leak rate of Ca\textsuperscript{2+} from ER & 0.15 s\textsuperscript{-1} & This work \\
        $V_{max}^\mathrm{MDH}$ & Limiting rate of MDH & 32 mM s\textsuperscript{-1} & This work \\
        $V^\mathrm{NCX}_{max}$ & Limiting rate of Na\textsuperscript{+}/Ca\textsuperscript{2+} exchanger & $2 \times 10^{-3}$ mM s\textsuperscript{-1} & This work \\
        $V_{max}^\mathrm{SDH}$ & Limiting rate of SDH & 1 mM s\textsuperscript{-1} & This work \\
        $V^\mathrm{SERCA}_{max}$ & Limiting rate of SERCA pumps & 0.12 mM s\textsuperscript{-1} & \cite{wacquier_interplay_2016} \\
        $V_{max}^\mathrm{UNI}$ & Limiting rate of mitochondrial uniporter & 0.30 mM s\textsuperscript{-1} & This work \\
         & & & \\
        \hline
    \end{tabular}
    }
    \end{fullwidth}
\end{table}


\section{Acknowledgments}

VV is funded by the Complex Living Systems Initiative at the University of Luxembourg. FA and ME are funded by the Luxembourg National Research Fund, grant ChemComplex (C21/MS/16356329). GF is funded by the European Union -- NextGenerationEU -- and by the program STARS@UNIPD with project ``ThermoComplex''. FA, AS and ME acknowledge financial support of the Institute for Advanced Studies of the University of Luxembourg through an Audacity Grant (IDAE-2020).


\bibliography{bibThermoMitoCa}



\end{document}